\documentclass{new_tlp}
\usepackage{mathptmx}
\usepackage{ifpdf}

\newif\ifextended\extendedfalse

\extendedtrue

\newif\ifdraft\draftfalse
\newif\iffinal\finaltrue
\newif\ifdotikz\dotikztrue
\newif\ifinlineref\inlinereffalse

\dotikztrue
\newcommand\bcmdtab{\noindent\bgroup\tabcolsep=0pt%
  \begin{tabular}{@{}p{10pc}@{}p{20pc}@{}}}
\newcommand\ecmdtab{\end{tabular}\egroup}

  \title[Theory and Practice of Logic Programming]
        {Ticker: A System for Incremental ASP-based Stream Reasoning\thanks{This research has been supported by the Austrian Science Fund (FWF) projects P26471 and W1255-N23.}}

  \author[H. Beck, T. Eiter, C. Folie]
    {Harald Beck, Thomas Eiter, and Christian Folie\\
    Institute of Information Systems, Vienna University of Technology\\ 
  Favoritenstra\ss{}e\ 9-11, A-1040 Vienna, Austria\\ 
  \email{$\{$beck,eiter$\}$@kr.tuwien.ac.at, christian.folie@outlook.com}}

\jdate{July 2017} 
\pubyear{2017} 
\pagerange{\pageref{firstpage}--\pageref{lastpage}}

\newtheorem{lemma}{Lemma}[section]

\usepackage[lined,linesnumbered,algoruled]{algorithm2e}
\usepackage{algpseudocode}
\usepackage{caption}

\usepackage[utf8]{inputenc}

\usepackage{graphicx}
\graphicspath{{figures/}}
\usepackage{latexsym}
\usepackage{amsmath}
\usepackage{amstext}
\usepackage{amssymb}
\usepackage{amsfonts}
\usepackage{color}
\usepackage{paralist}
\usepackage{enumerate}
\usepackage{ifpdf}
\usepackage{marvosym}
\usepackage{float}
\usepackage{url}
\usepackage{multirow}
\usepackage{alltt}

\usepackage[amsmath,thmmarks]{ntheorem}

\newcommand{\nop}[1]{}

\newcommand{\reduceItemSkip}{\setlength{\itemsep}{-0.5ex}}

\newcommand{\medblacksquare}{\mathbin{\scalebox{0.6}{\ensuremath{\blacksquare}}}} 

\newcommand{\leanparagraph}[1]{\smallskip\noindent\textbf{#1.} }

\newcounter{myenumctr}
\newenvironment{myenumerate}{\begin{list}{(\arabic{myenumctr})}{\usecounter{myenumctr}
\topsep=0pt
\setlength{\leftmargin}{0pt}
\setlength{\itemindent}{\labelwidth}
\setlength{\itemsep}{0cm}}}
{\end{list}}

\newcommand{\mi}[1]{\ensuremath{\mathit{#1}}}

\newcommand{\intensional}{I}
\newcommand{\extensional}{E}

\newcommand{\Atoms}{A}

\renewcommand{\vec}[1]{{\mathbf{#1}}}

\newcommand{\tup}[1]{\ensuremath{\langle #1 \rangle}}

\newcommand{\backgroundData}{\mathrm{B}}

\newcommand{\head}{\ensuremath{\mathit{H}}}
\newcommand{\body}{\ensuremath{\mathit{B}}}

\newcommand{\ExtendedAtoms}{\ensuremath{\Atoms^+}}

\newcommand{\opPlaceholder}{\ensuremath{\mathbin{\star}}}

\def\wsExThree{3.0}

\mathchardef\minus="2D

\makeatletter
\newcommand{\dotminus}{\mathbin{\text{\@dotminus}}}
\newcommand{\@dotminus}{%
  \ooalign{\hidewidth\raise1ex\hbox{.}\hidewidth\cr$\m@th-$\cr}%
}
\makeatother

\newcommand{\intpr}{\upsilon} 
\newcommand{\tickIntpr}{v}

\newcommand{\tickStream}{\dot{S}}

\def\AS{AS}
\DeclareMathOperator{\naf}{not}

\newcommand{\wfnTime}{\tau}
\newcommand{\wfnTuple}{\#}

\newcommand{\window}{\ensuremath{\boxplus}}
\newcommand{\timeWindow}[1]{\ensuremath{\boxplus}^{#1}}
\newcommand{\tupleWindow}[1]{\ensuremath{\boxplus}^{\# #1}}

\SetKwProg{Def}{def}{$\,$:}{}
\SetKwProg{Defn}{defn}{~$=$}{}
\SetKwProg{DefnCustom}{defn}{}{}
\SetKw{Let}{let}
\SetKw{Halt}{halt}
\SetKwInput{KwIn}{Input}
\SetKwFor{ForEach}{foreach}{}{}
\SetKw{Or}{or}
\SetKw{And}{and}
\SetKwIF{If}{ElseIf}{Else}{if}{}{else if}{else}{endif}
\SetKw{Match}{match}
\SetKw{Case}{case}
\SetKw{MapTo}{\ensuremath{\;\;\Rightarrow\;\;}}

\newcommand{\assign}{\ensuremath{~\mathtt{:=}~}}
\newcommand{\algocomment}[1]{\text{//$\,$#1}}

\newcommand{\blockskip}{\smallskip}

\newcommand{\enc}[1]{\hat{#1}}

\newcommand{\alphaval}{\ensuremath{\mathit{alpha}}}
\newcommand{\alphahigh}{\ensuremath{\mi{high}}}
\newcommand{\alphamid}{\ensuremath{\mi{mid}}}
\newcommand{\alphalow}{\ensuremath{\mi{low}}}
\newcommand{\lfu}{\ensuremath{\mathit{lfu}}}
\newcommand{\lru}{\ensuremath{\mathit{lru}}}
\newcommand{\fifo}{\ensuremath{\mathit{fifo}}}
\newcommand{\done}{\ensuremath{\mathit{done}}}
\newcommand{\random}{\ensuremath{\mathit{random}}}

\newcommand{\cache}{\mathit{cache}}
\newcommand{\avail}{\mathit{avail}}
\newcommand{\itemm}{\mathit{item}}
\newcommand{\node}{\mathit{node}}
\newcommand{\req}{\mathit{req}}
\newcommand{\getFrom}{\mathit{get}}
\newcommand{\nGetFrom}{\mathit{nGet}}
\newcommand{\source}{\mathit{source}}
\newcommand{\qual}{\mathit{qual}}
\newcommand{\level}{\mathit{lev}}
\newcommand{\down}{\mathit{down}}
\newcommand{\qLev}{\mathit{qLev}}
\newcommand{\need}{\mathit{need}}
\newcommand{\reach}{\mathit{reach}}
\newcommand{\conn}{\mathit{conn}}
\newcommand{\edge}{\mathit{edge}}

\newcommand{\bbN}{\ensuremath{\mathbb{N}}}

\let\oldnl\nl
\newcommand{\nonl}{\renewcommand{\nl}{\let\nl\oldnl}}

\newcommand{\timePinned}[1]{#1_@}
\newcommand{\tickPinned}[1]{#1_\#}

\newcommand{\now}{\mathit{now}}
\newcommand{\cnt}{\mathit{cnt}}
\newcommand{\tick}{\mathit{tick}}

\newcommand{\encWindowAtom}{\omega}

\newcommand{\signalSet}{\mathit{Sig}}

\newcommand{\tTick}{\ensuremath{t_{\mathit{tick}}}}
\newcommand{\tInit}{\ensuremath{t_{\mathit{init}}}}
\newcommand{\tTotal}{\ensuremath{t_{\mathit{total}}}}

\newtheorem{theorem}{Theorem}
\newtheorem{proposition}{Proposition}
\renewtheorem{lemma}{Lemma}

\newtheorem{definition}{Definition}

\theorembodyfont{\normalfont}
\newcounter{examplecounter}
\theoremsymbol{\ensuremath{\medblacksquare}}
\newtheorem{example}[examplecounter]{Example}
\theorembodyfont{\normalfont}
\theoremheaderfont{\normalfont\itshape}
\theoremseparator{.~}
\theoremsymbol{\ensuremath{\Box}}

\ifdotikz
\usepackage{pgf}
\usepackage{tikz}
\usetikzlibrary{arrows,positioning,automata,decorations,fit,backgrounds,calc}
\usepgflibrary{shapes.geometric} %
\usetikzlibrary{shapes.geometric} %
\usepgflibrary{decorations.pathmorphing} %
\usetikzlibrary{decorations.pathmorphing} %
\usepgflibrary{decorations.text} %
\usetikzlibrary{decorations.text} %
\usetikzlibrary{matrix}
\pgfrealjobname{ms}
\else
\ifpdf
\long\def\beginpgfgraphicnamed#1#2\endpgfgraphicnamed{\includegraphics{#1}}
\else
\usepackage{epsfig}
\long\def\beginpgfgraphicnamed#1#2\endpgfgraphicnamed{\epsfig{file=#1.eps}}
\fi
\fi

\draftfalse

\begin{document}

\label{firstpage}

\maketitle

\begin{abstract}
In complex reasoning tasks, as expressible by Answer Set Programming
(ASP), problems often permit for multiple solutions. In dynamic
environments, where knowledge is continuously changing, the question
arises how a given model can be incrementally adjusted relative to new
and outdated information.
This paper introduces Ticker, a prototypical engine for well-defined
logical reasoning over streaming data. Ticker builds on a practical
fragment of the recent rule-based language LARS which extends Answer
Set Programming for streams by providing flexible expiration control
and temporal modalities.
We discuss Ticker's reasoning strategies: First, the repeated one-shot
solving mode calls Clingo on an ASP encoding. We show how this
translation can be incrementally updated when new data is streaming in
or time passes by. Based on this, we build on Doyle's classic
justification-based truth maintenance system (TMS) to update models of
non-stratified programs. %
Finally, we empirically compare the obtained evaluation mechanisms.
This paper is under consideration for acceptance in TPLP.
\end{abstract}

\begin{keywords}
Stream Reasoning, Answer Set Programming, Nonmonotonic Reasoning
\end{keywords}

\section{Introduction}\label{sec:introduction}

Stream reasoning~\cite{VCHF09} as research field emerged from data
processing \cite{BabuW01}, i.e., the handling of continuous queries in a
frequently changing database. Work in Knowledge Representation \&
Reasoning, e.g.~\cite{RenP11,ricochet}, shifts the focus from high throughput
to high expressiveness of declarative queries and programs. In
particular, the logic-based framework LARS~\cite{bdef15lars} was
defined as an extension of Answer Set Programming (ASP) with window
operators for deliberately dropping data, e.g., based on time or counting atoms,
and controlling the temporal modality in the resulting windows.

When dealing with complex reasoning tasks in stream settings, one may in
general not afford to recompute models from scratch every time new data
comes in or when older portions of data become outdated. Besides the
pragmatic need for efficient computation, there is also a semantic issue:
while aspects of a solution might have to change dynamically and
potentially quickly, typically not everything should be reconstructed
from scratch, but adapted to fit the current data.

Recently, many stream processing tools and reasoning features have been
proposed, e.g.~\cite{Barbieri10,PhuocDPH11,clingo14}. However, an
ASP-based stream reasoning engine that supports window operators and has an
incremental model update mechanism is lacking to date. This may be
explained by the fact that nonmonotonic negation, beyond recursion, makes efficient
incremental update non-trivial; combined with temporal reasoning
modalities over data windows, this becomes even more challenging.

\leanparagraph{Contributions} We tackle this issue and make the
following contributions.
\begin{myenumerate}
\item We present a notion of tick streams to formally represent the
  sequential steps of a fully incremental stream reasoning system.
\item Based on this, we give an intuitive translation of a practical
  fragment of LARS programs, plain LARS, to ASP
  suitable for standard one-shot solving, and in particular, stratified programs.
\item Next, we develop an ASP encoding that can be incrementally updated
  when time passes by or when new input arrives.
\item We then present Ticker, our prototype reasoning engine that comes
  with two reasoning strategies. One utilizes Clingo~\cite{clingo14}
  with a static ASP encoding, the other
  truth maintenance techniques~\cite{Doyle79} to adjust models based on the
  incremental encoding.
\item Finally, we 
experimentally compare the two reasoning modes in application scenarios.
The results demonstrate the performance benefits that arise from incremental
evaluation.
\end{myenumerate}
\smallskip
In summary, we provide a novel technique for adjusting an ASP-based
stream reasoning program by time and data streaming in. In particular,
the update technique of the program is independent of the model update
technique used to process the program change.

\section{Stream Reasoning in LARS}\label{sec:lars}
We will gradually introduce the central
concepts of LARS \cite{bdef15lars} tailored to the considered fragment. If appropriate, we give
only informal descriptions.

Throughout, we distinguish \emph{extensional atoms} $\Atoms^\extensional$ for input data and \emph{intensional atoms} $\Atoms^\intensional$ for derived information. By $\Atoms = \Atoms^\extensional \cup \Atoms^\intensional$, we denote the set of~\emph{atoms}.
\begin{definition}[Stream]
 \label{def:stream}
  A stream ${S=(T,\intpr)}$ consists of a \emph{timeline} $T$, which is
 a closed 
 nonempty interval in $\bbN$, and an \emph{evaluation
    function} ${\intpr : \bbN \mapsto 2^\Atoms}$. The elements ${t \in T}$
  are called \emph{time points}.
\end{definition}
Intuitively, a stream $S$ associates with each time point a set of
atoms.  We call $S$ a \emph{data stream}, if it contains only
extensional atoms.
To cope with the amount of data, one usually considers only
recent atoms.
Let ${S=(T,\intpr)}$ and ${S'=(T',\intpr')}$ be two streams such that ${S'
\subseteq S}$, i.e., ${T' \subseteq T}$ and ${\intpr'(t') \subseteq
  \intpr(t')}$ for all ${t' \in T'}$. Then $S'$ is called a
\emph{window} of $S$.
\begin{definition}[Window function]\label{def:window-function}
  Any (computable) function $w$ that returns, given a stream
  $S=(T,\intpr)$ and a time point ${t \in \bbN}$, a \emph{window} $S'$
  of $S$, is called a \emph{window function}.
\end{definition}
Widely used are \emph{time-based} window functions, which select all
atoms appearing in last $n$ time points, and \emph{tuple-based} window
functions, which select a fixed number of latest tuples.
To this end, we define the \emph{tuple size $|S|$} of a stream $S=(T,\intpr)$ as
$|\{ (a,t) \mid t \in T, a \in \intpr(t)\}|$.

\begin{figure}[t]
  \centering
  \footnotesize
  \beginpgfgraphicnamed{buses-pic0}  
  \begin{tikzpicture}[scale=1.20,node distance=0.4cm,>=latex]
      \draw [->] (34.5,0) -- ($(40,0)+(2,0)$) node [anchor=west] {};
 
      \foreach \i in {35,36,37,38,39,40,41} 
      {
        \draw (\i cm,0pt)  node[anchor=north] at (\i,-0.32)
        {$\i$};
        \node at (\i,0) {\scriptsize $\bullet$};
      }
      \foreach \i in {35,37,39} 
      {
        \draw (\i cm,0pt)  node[anchor=north] at (\i,-0.32)
        {$\i$};
        \draw [dotted] (\i,0) -- (\i,0.4) ;
      }      

      \draw [thick] ($(40,0.3)$) -- ($(40,-0.3)$) ;

      \node (first-a) at  ($(35,0.55)$) {$\{a(x)\}$};
      \node (busG) at  ($(37,0.55)$) {$\{a(y),a(z)\}$};
      \def\closeLabelsOffset{0.0}
      \node (tramB) at  ($(39,0.55)-(\closeLabelsOffset,0)$) {$\{a(x)\}$};

      \def\windowHeight{0.16}      
      \draw [thick] (40.0,\windowHeight) -- ($(40.0,\windowHeight)-(\wsExThree,0)$) -- ($(40.0,-\windowHeight)-(\wsExThree,0)$) -- (40.0,-\windowHeight);
    \end{tikzpicture}
  \endpgfgraphicnamed
  \caption{Temporal extent of a sliding tuple-based window of size 3 (or 2) at $t=40$}
  \label{fig:window}
\end{figure}
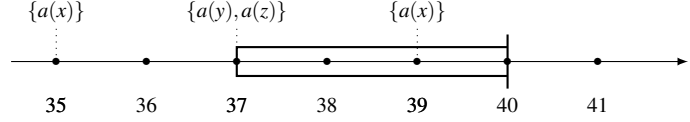

\begin{definition}[Sliding Time-based and Tuple-based Window]
  \label{def:sliding-lars-windows}
  Let ${S=(T,\intpr)}$ be a stream, $t \in T=[t_1,t_m]$ and let
  ${n \in \bbN \cup \{\infty\}}$. Then, 
  
\begin{itemize}
\item[(i)] the \emph{sliding time-based window function $\wfnTime_n$ (for
    size~$n$)} is ${ \wfnTime_n(S,t) = (T',\intpr|_{T'})}$, 
    where ${T'=[t',t]}$ and ${t' = \max\{t_1,t-n\}}$;
\item[(ii)] the \emph{sliding tuple-based window function
    $\wfnTuple_n$ (for size~$n$)} is   %
  \begin{displaymath}
    \wfnTuple_n(S,t) =
    \begin{cases}
      \wfnTime_{t-t'}(S,t) & \text{if}~|\wfnTime_{t-t'}(S,t)| \leq n,\\
      S' & \text{else,}
    \end{cases}
  \end{displaymath}
where   $t' = \max (\{u \in T \mid |\wfnTime_{t-u}(S,t)| \geq n\} \cup \{t_1\})$
and $S'=([t',t],\intpr')$ has tuple size $|S'|=n$ such that 
$\intpr'(u)=\intpr(u)$  for all $u \in [t'+1,t]$ and
  $\intpr'(t')\subseteq \intpr(t')$.
\end{itemize}
\end{definition}
Note that in general,  multiple options exist for defining $\intpr'$
at $t'$ in the tuple-based window. However, we assume a deterministic
choice as specified by the implementation of the function. In
particular, we will later consider that atoms are streaming in an order,
which leads to a natural, unique cut-off position based on counting.
\begin{example}\label{ex:window}
  Fig.~\ref{fig:window} window depicts at partial stream
  $S=([35,41],\intpr)$, where $\intpr= \{ 35 \mapsto \{a(x)\},$
  $37 \mapsto \{a(y),a(z)\}, 39 \mapsto \{a(x)\}\}$, and a time
  window of length $3$ at time $t=40$, which corresponds to a
  tuple window of size $3$ there. Notably, there are two options
  for a tuple window of size $2$, both of which select timeline
  $[37,40]$, but only one of the atoms at time $37$, respectively.~
\end{example}
We also use window functions %
with streams as single argument, applied implicitly at the end of the
timeline, i.e., if $S=([t_0,t],\intpr)$, then $\wfnTime_n(S)$
abbreviates $\wfnTime_n(S,t)$ and $\wfnTuple_n(S)$ stands for
$\wfnTuple_n(S,t)$.

\leanparagraph{Window operators $\window^w$}
A window function $w$ can be accessed in rules by window operators.
That is to say, an expression $\window^w \alpha$ has the effect that
$\alpha$ is evaluated on the ``snapshot'' of the data stream delivered
by its associated window function $w$. Within the selected snapshot,
LARS allows for controlling the temporal semantics with further modalities.

\leanparagraph{Temporal modalities}
Let ${S=(T,\intpr)}$ be a stream,
${a \in \Atoms}$ and ${\backgroundData \subseteq \Atoms}$ static \emph{background data}.
Then, at time point ${t \in T}$,
\begin{itemize}\reduceItemSkip
\item $a$ holds, if ${a \in \intpr(t)}$ or ${a \in \backgroundData}$;
\item $\Diamond a$ holds, if $a$ holds at some time point ${t' \in T}$;
\item $\Box a$ holds, if $a$ holds at all time points ${t' \in T}$; and
\item $@_{t'} a$ holds, where $t' \in \bbN$, if $t' \in T$ and $a$ holds at $t'$.
\end{itemize}
The set $\ExtendedAtoms$ of \emph{extended atoms} $e$ is given by the grammar
$
e\; ::=\;   a \mid @_t a \mid \window^w @_t a \mid \window^w \Diamond a \mid \window^w \Box a\,,
$
where $a \in \Atoms$ and $t$ is any time point.  The expressions 
$@_ta$ are called \emph{$@$-atoms};
${\window^w \opPlaceholder a}$, where
${\opPlaceholder \in \{@_t, \Diamond, \Box\}}$, are
\emph{window atoms}. We write $\timeWindow{n}$ for $\window^{\wfnTime_n}$,
which is not to be confused with  $\tupleWindow{n}$.

\begin{example}[cont'd]
At $t=40$,
$\timeWindow{3}\Diamond a(x)$ and $\timeWindow{3}@_{37}a(y)$ hold, as
does $\tupleWindow{1} \Box a(x)$ at $t=35,39$.~
\end{example}
\subsection{Plain LARS Programs}\label{sec:plain-lars}

We use a fragment of the formalism in~\cite{bdef15lars}, called
\emph{plain LARS} programs.

\leanparagraph{Syntax}
A \emph{(ground plain LARS) program} $P$ is a set of rules
of the form
\begin{equation}\label{eq:rule-syntax}
  \alpha \leftarrow \beta_1,\dots,\beta_j, \naf \beta_{j+1},\dots, \naf
\beta_{n}\,,
\end{equation}
where the \emph{head} $\alpha$ is of form $a$ or
$@_t a$, ${a \in \Atoms^\intensional}$, and in the 
\emph{body} $\beta(r)=\beta_1,\dots,\beta_j, \naf \beta_{j+1},\dots, \naf
\beta_{n}$ each $\beta_i$ is an extended atom.
We let 
$H(r)=\alpha$ and $\body(r)=\body^+(r) \cup \body^-(r)$,
where $\body^+(r)=\{\beta_1,$ $\ldots,$ $\beta_j\}$ and 
$\body^-(r)=\{\beta_{j+1},\ldots,\beta_n\}$ are the
the \emph{positive}, resp. \emph{negative body atoms} of $r$.

\leanparagraph{Semantics}
For a data stream ${D=(T_D,\intpr_D)}$, 
any stream ${I=(T,\intpr) \supseteq D}$ that coincides with $D$ on 
$\Atoms^\extensional$, i.e., $a {\,\in\,} \intpr(t) \cap \Atoms^\extensional$ iff $a {\,\in\,} \intpr_D(t)$, is an \emph{interpretation
stream}\/ for $D$.
A tuple ${M=\langle I, W,\backgroundData\rangle}$, where $W$ is a set of
window functions and $\backgroundData$ is the background
knowledge, %
is then an \emph{interpretation} for $D$. Throughout, we assume
$W=\{\wfnTime_k, \wfnTuple_n \mid k,n \in \bbN\}$ and $\backgroundData$
are fixed and %
also
omit them.

Satisfaction by $M$ at ${t \in T}$ is as follows: ${M,t \models
\alpha}$ for ${\alpha \in \ExtendedAtoms}$, if $\alpha$ holds in
${(T,\intpr)}$ at time ${t}$; 
${M,t \models r}$ for rule $r$, if ${M,t \models \beta(r)}$ implies
${M,t \models H(r)}$, where ${M,t \models \beta(r)}$, if 
\begin{inparaenum}[(i)]
\item ${M,t \models \beta_i}$ for all ${i \in \{1,\dots,j\}}$ and
\item ${M,t \not\models \beta_i}$ for all ${i \in \{j{+}1,\dots,n\}}$; and
\end{inparaenum}
${M,t \models P}$ for  program $P$, i.e., $M$ is a \emph{model} of $P$ (for $D$) at
${t}$, if ${M,t \models r}$ for all ${r \in P }$.
Moreover, ${M}$ is \emph{minimal}, 
if in addition no model
${M'=\langle S',W,\backgroundData \rangle\neq M}$ of $P$ exists
such that $S'=(T,\intpr')$ and ${\intpr' \subseteq \intpr}$.
\begin{definition}[Answer Stream]
  An interpretation stream $I$ is an answer stream of program $P$ for
  the data stream ${D \subseteq I}$ at time $t$, if
  ${M=\langle I,W,\backgroundData \rangle}$ is a minimal model of the
  \emph{reduct} $P^{M,t}=\{r \in P \mid M,t \models \beta(r)\}$. By
  $\AS(P,D,t)$ we denote the set of all such answer streams $I$.
\end{definition}
\begin{example}[cont'd]
  Consider $D$ from Fig.~\ref{ex:window} and
  $P=\{ b(x) \leftarrow \timeWindow{3} \Diamond a(x) \}$. Then, for all
  $t\in [35,41]$ the answer stream $I$ at $t$ is unique and adds to $D$
  the mapping $t \mapsto \{b(x)\}$.
\end{example}
\leanparagraph{Non-ground programs} The semantics for LARS is formally
defined for ground programs but extends naturally for the non-ground
case by considering the respective ground instantiations.

\leanparagraph{Windows on intensional/extensional atoms}
For practical reasons, we consider tuple windows only on extensional
data. Their intended use is counting input data, not inferences;
using them on intensional data is conceptually questionable.
\begin{example}\label{ex:tuple-window-extensional}
  Consider the rule $r = b \leftarrow \tupleWindow{1}\Diamond a$ and the
  stream $S=([0,1],\{0 \mapsto \{a\}\})$, which is not a model for $r$,
  since the rule fires and we thus must have $b$ at time $1$. However,
  in this interpretation, $\tupleWindow{1}\Diamond a$ does not hold any
  more, if we also take into account the inference $b$. Thus, the
  interpretation would not be minimal. Moreover, further inferences
  would not be founded. Hence, program $\{r\}$ has no model.
\end{example}
In contrast to tuple windows, time windows are useful and allowed on
arbitrary data, as long as no cyclic positive dependencies through
time-based window atoms $\window^n\Box a$ occur.
\begin{example}\label{ex:alpha-high-abstraction}
  Assume a range of values $V=0,\dots,30$, among which $V\geq 18$ are
  considered `high.' To test whether the predicate $\mi{alpha}$ always had a
  high value during the last $n$ time points, we first abstract by
  $@_T \alphahigh \leftarrow \timeWindow{n} @_T \mi{alpha}(V), V\geq 18$ for
  and then test $\mi{yes} \leftarrow \timeWindow{n} \Box \alphahigh$.~
\end{example}

\section{Static ASP Encoding}
\label{sec:lars-to-asp}

In this section we will first give a translation of LARS programs $P$ to
an ASP program $\enc{P}$. Toward incremental evaluation of $P$, we will
then show how $\enc{P}$ can be adjusted to accommodate new input signals
and account for expiring information as specified by window operators.

\begin{definition}[Tick]\label{def:tick}
  A pair $k=(t,c)$, where $t,c \in \bbN$, is called a \emph{tick},
  with~$t$ the \emph{(tick) time} and~$c$ the \emph{(tick) count};
  $(t+1,c)$ is called the \emph{time increment} and $(t,c+1)$ the
  \emph{count increment} of $k$. A sequence
  $K=\langle k_1,\dots,k_m \rangle$, $m\geq 1$, of ticks is a \emph{tick
    pattern}, if every tick $k_{i+1}$ is either a time increment or a
  count increment of $k_i$.
\end{definition}
Intuitively, a tick pattern captures the incremental development of a
stream in terms of time and tuple count, where at each step exactly one
dimension increases by 1. For a set of ticks, at most one linear
ordering yields a tick pattern. Thus, we can view a tick pattern $K$
also as set.
\begin{definition}[Tick Stream]\label{def:tick-stream}
  A \emph{tick stream} is a pair $\tickStream=(K,\tickIntpr)$
of a tick pattern $K$ and an \emph{evaluation function}
$\tickIntpr$ s.t. $\tickIntpr(k_{i+1}) = \{a\}$ for some $a \in \Atoms$,
  if $k_{i+1}$ is a count increment of $k_i$, else
  $\tickIntpr(k_{i+1})=\emptyset$.
\end{definition}
We say that a tick stream ${\tickStream=(K,\tickIntpr)}$ 
with $K=\langle (t_1,c_1),\dots,(t_m,c_m) \rangle$
is \emph{at tick} $(t_m,c_m)$.
By default, we 
assume $(t_1,c_1)=(0,0)$ and thus $c_m$ is the total
number of atoms.
We also write $\tickIntpr(t,c)$ instead of $\tickIntpr((t,c))$. %
Naturally, a \emph{(tick) substream}
$\tickStream' \subseteq \tickStream$ is a tick stream
$(K',\tickIntpr')$, where $K'$ is a subsequence of $K$ and $\tickIntpr'$
is the restriction $\tickIntpr|_{K'}$ of $\tickIntpr$ to $K'$, i.e.,
$\tickIntpr'(t,c) = \tickIntpr(t,c)$ if $(t,c) \in K'$, else
$\tickIntpr'(t,c)=\emptyset$.

\begin{example}\label{ex:tick-stream}
  The sequence
  $K=\langle (0,0),(1,0),(2,0),(3,0),(3,1),(3,2),(4,2) \rangle$ is a
  ``canonical''
  tick pattern starting at $(0,0)$, where $(3,1)$ and $(3,2)$ are the only count
  increments. Employing an evaluation $\tickIntpr(3,1)=\{a\}$ and
  $\tickIntpr(3,2)=\{b\}$, we get a tick stream
  $\tickStream=(K,\tickIntpr)$ which is at tick $(4,2)$.
\end{example}

\begin{definition}[Ordering]\label{def:ordering}
  Let $\tickStream=(K,\tickIntpr)$ be a tick stream, where
  $K=\langle (t_1,c_1),\dots,(t_m,c_m)\rangle$, and let $S=(T,\intpr)$ be a
  stream such that $T=[t_1,t_m]$ and
  $\intpr(t)= \bigcup \{ \tickIntpr(t,c) \mid (t,c) \in K\}$ for all
  $t \in T$. Then, we say $\tickStream$ is an \emph{ordering} of $S$,
  and $S$ \emph{underlies} $\tickStream$.
\end{definition}
Note that in general, a stream $S$ has multiple orderings, but every
tick stream $\tickStream$ has a unique underlying stream. All
orderings of a stream have the same tick pattern.
\begin{example}[cont'd]\label{ex:ordering}
  Stream $S=([0,4],\intpr)$, where $\intpr =\{3 \mapsto \{a,b\}\}$, is
  the underlying stream of $\tickStream$ of
  Ex.~\ref{ex:tick-stream}. 
  A further ordering of $S$ is
  $\tickStream'=(K,\tickIntpr')$, where
  $\tickIntpr'=\{ (3,1) \mapsto \{b\}, (3,2) \mapsto \{a\}\}$.~
\end{example}
Sliding windows as
in Def.~\ref{def:sliding-lars-windows}
carry over naturally for tick streams. There are two central
differences. First, ticks replace time points as positions in a stream,
and thus as second argument of the window functions. Second, tuple-based
windows are now always unique.
\begin{definition}[Sliding Windows over Tick Streams]\label{def:sliding-windows-over-tick-streams}
  Let~$\tickStream=(K,\tickIntpr)$ be a tick stream, where
  $K=\langle (t_1,c_1),\dots,(t_m,c_m)\rangle$ and $(t,c) \in K$. Then
  the \emph{time window function $\wfnTime_n$,
    $n \geq 0$}, is defined by
  $\wfnTime_n(\tickStream,(t,c)) = (K',\tickIntpr|_{K'})$, where
  $K' = \{(t',c') \in K \mid \max\{t_1,t-n\} \leq t' \leq t\}$, and the
  \emph{tuple window function $\wfnTuple_n$, $n \geq 1$},
  by $\wfnTuple_n(\tickStream,(t,c)){=}(K',\tickIntpr|_{K'})$, where
  $K'{=}\{ (t',c') \in K {\,\mid} \max\{c_1,c-n+1\} {\,\leq\,} c' {\,\leq\,} c\}$.
\end{definition}
As for Def.~\ref{def:sliding-lars-windows}, we consider windows
over tick streams also implicitly at the end of the
timeline.

\begin{lemma}\label{lemma:time-window-underlies} If stream $S$ underlies tick stream $\tickStream$, then
  $\wfnTime_n(S)$ underlies $\wfnTime_n(\tickStream)$.
\end{lemma}
\begin{example}[cont'd]
  Given $\tickStream$ and $S$ from Example~\ref{ex:ordering}, we have
  $\wfnTime_1(\tickStream,4)=(\langle (3,0),(3,1),(3,2),$ $(4,2) \rangle,\tickIntpr)$ with underlying stream
  $\wfnTime_1(S,4)=([3,4],\intpr)$.
\end{example}
Correspondence for tuple windows is more subtle due to the different
options to realize them.
\begin{lemma}\label{lemma:tuple-window-underlies}
  Let stream $S$ underlie tick stream $\tickStream$ and assume the tuple
  window $\#_n(S)$ is based on the order in which atoms appeared in
  $S$. Then, $\wfnTuple_n(S)$ underlies $\wfnTuple_n(\tickStream)$.
\end{lemma}
\begin{example}[cont'd]
  Stream $S$ has two tuple windows of size $1$:
  $S_a=([3,4],\{3 \mapsto \{a\}\})$ and
  $S_b=([3,4],\{3 \mapsto \{b\}\})$; the latter underlies
  $\wfnTuple_1(\tickStream)=(\langle (3,2),(4,2) \rangle, (3,2) \mapsto
  \{b\})$.~
\end{example}
We can represent a stream~$S=(T,\intpr)$ alternatively by $T$ and a set
of \emph{time-pinned} atoms, i.e., the set
$\{ \timePinned{a}(\vec{x},t) \mid a(\vec{x}) \in
\intpr(t), t \in T\}$. Similarly, tick streams can be modelled by
\emph{tick-pinned} atoms of form $\tickPinned{a}(\vec{x},t,c)$,
where $c$ increases by 1 for every incoming signal.
\begin{example}[cont'd]
  Given extra knowledge about the time $t=4$, stream $S$ is fully
  represented by $\{\timePinned{a}(3), \timePinned{b}(3)\}$, whereas tick stream
  $\tickStream$ can be encoded by the set
  $\{ \tickPinned{a}(3,1), \tickPinned{b}(3,2)\}$.
\end{example}
The notions of data/interpretation stream readily carry over to their
tick analogues.
Moreover, we say a
tick interpretation stream $I$ is an \emph{answer stream} of
program $P$ (for tick data stream $D$ at $t$), if the
underlying stream $I'$ of $I$ is an answer stream of $P$ (for the
underlying data stream $D'$ at $t$).
\begin{algorithm}[t]
  \caption{Plain LARS Program to ASP $\mi{LarsToAsp}(P,t)$}
  \label{alg:lars-program-to-asp}
  \DontPrintSemicolon
  \SetAlgoVlined
  \small

  \KwIn{A (potentially non-ground) plain LARS program $P$,
    and the evaluation time point $t$}

  \KwOut{ASP encoding $\enc{P}$, i.e., a set of normal logic rules}

  \blockskip

  $Q \assign \{\, a(\vec{X}) \leftarrow \mi{now}(\dot{N}), \timePinned{a}(\vec{X},\dot{N});\; \timePinned{a}(\vec{X},\dot{N}) \leftarrow \mi{now}(\dot{N}), a(\vec{X}) \mid a \text{ is a predicate in }P\}$\label{alg-line:lars-program-to-asp:Q}\;
  $R \assign \bigcup_{r \in P} \mi{larsToAspRules}(r)$\;
  \Return $Q \cup R \cup \{\mi{now(t)} \}$

  \blockskip
  \hrule
  \blockskip

  \lDefn{$\mi{larsToAspRules}(r)$}{$\{ \mi{baseRule}(r)\} \cup\, \bigcup_{i=1}^m \mi{windowRules}(e_i)$}

  \blockskip

  \Defn{$\mi{baseRule}(h \leftarrow e_1,\dots,e_n, \naf e_{n+1},\dots,\naf e_{m})$}{
    $\mi{atm}(h) \leftarrow \mi{atm}(e_1),\dots,\mi{atm}(e_n), \naf\,
    \mi{atm}(e_{n+1}),\dots,\naf\, \mi{atm}(e_{m})$
  }

  \blockskip
  
  \DefnCustom{$\mi{atm}(e)=\Match~e$}{
      \Case~~$a(\vec{X}) \MapTo a(\vec{X})$\label{alg-line:atm:start}\\
      \Case~~$@_{T} a(\vec{X}) \MapTo \timePinned{a}(\vec{X},T)$\\
      \Case~~$\window^w @_{T} a(\vec{X}) \MapTo \encWindowAtom_e(\vec{X},T)$\quad\algocomment{$\encWindowAtom_e$ is a fresh predicate associated with $e$}\\
      \Case~~$\window^w \Diamond a(\vec{X}) \MapTo \encWindowAtom_e(\vec{X})$\\
      \Case~~$\window^w \Box a(\vec{X}) \MapTo \encWindowAtom_e(\vec{X})$\\
   \label{alg-line:atm:end}}
  
   \blockskip

   \DefnCustom{$\mi{windowRules}(e)=\Match~e$}{
      \Case~~$\timeWindow{n} @_{T} a(\vec{X}) \MapTo \{\, \encWindowAtom_e(\vec{X},T) \leftarrow \mi{now}(\dot{N}), \timePinned{a}(\vec{X},T), T=\dot{N}-i \mid i=0,\dots,n\,\}$\label{alg-line:rules:start}\\
      \Case~~$\timeWindow{n} \Diamond a(\vec{X}) \MapTo \{\, \encWindowAtom_e(\vec{X}) \leftarrow \mi{now}(\dot{N}), \timePinned{a}(\vec{X},T), T=\dot{N}-i \mid i=0,\dots,n\,\}$\\
      \Case~~$\timeWindow{n} \Box a(\vec{X}) \MapTo\{\, \encWindowAtom_e(\vec{X}) \leftarrow a(\vec{X}),\naf\, \mi{spoil}_e(\vec{X})\,\}\, \cup$\label{alg-line:window-rules:time-box-static}\\
      \quad$\{\, \mi{spoil}_e(\vec{X}) \leftarrow a(\vec{X}), \mi{now}(\dot{N}), \naf\, \timePinned{a}(\vec{X},T), T=\dot{N}-i \mid i=1,\dots,n\,\}$\label{alg-line:window-rules:time-box-spoil}\\
      %
      %
      %
      %
      \Case~~$\tupleWindow{n} @_{T} a(\vec{X}) \MapTo \{\, \encWindowAtom_e(\vec{X},T) \leftarrow \cnt(\dot{C}), \tickPinned{a}(\vec{X},T,D), D = \dot{C}-j \mid j=0,\dots,n-1\,\}$\label{alg-line:window-rules:tuple-at}\\
      \Case~~$\tupleWindow{n} \Diamond a(\vec{X}) \MapTo \{\, \encWindowAtom_e(\vec{X}) \leftarrow \cnt(\dot{C}), \tickPinned{a}(\vec{X},T,D), D = \dot{C}-j \mid j=0,\dots,n-1\,\}$\label{alg-line:window-rules:time-diamond}\\
      \Case~~$\tupleWindow{n} \Box a(\vec{X}) \MapTo\{\, \encWindowAtom_e(\vec{X}) \leftarrow a(\vec{X}),\naf\, \mi{spoil}_e(\vec{X})\,\}\,\cup$\label{alg-line:window-rules:tuple-box-static}\\
      \quad $\{\, \mi{spoil}_e(\vec{X}) \leftarrow a(\vec{X}), \cnt(\dot{C}), \tick(T,D), \dot{C}-n+1 \leq D \leq \dot{C},\, \naf\, \timePinned{a}(\vec{X},T)\,\}\,\cup$\label{alg-line:window-rules:tuple-box-spoil-time-range}\\
      \quad $\{\, \mi{spoil}_e(\vec{X}) \leftarrow a(\vec{X}), \cnt(\dot{C}), \tick(T,D), D=\dot{C}-n+1,\, \tickPinned{a}(\vec{X},T,D'),\, D'<D\,\}$\label{alg-line:window-rules:tuple-box-spoil-count-range}
      \\
      \lElse{$~\emptyset$ \label{alg-line:rules:end}}
  } 
  
\end{algorithm}

\leanparagraph{LARS to ASP (Algorithm~\ref{alg:lars-program-to-asp})}
Plain LARS programs extend normal logic programs by allowing extended
atoms in rule bodies, and also $@$-atoms in rule heads. Thus, if we
restrict $\alpha$ and $\beta_i$ in~\eqref{eq:rule-syntax} to atoms, we
obtain a normal rule. This observation is used for the translation of
LARS to ASP as shown in Algorithm~\ref{alg:lars-program-to-asp}.
The encoding has to take care of two central aspects. First, each
extended atoms $e$ is encoded by an (ordinary) atom $a$ 
that holds iff $e$ holds. Second, entailment in LARS is defined
with respect to some data stream $D$ and background data
$\backgroundData$ at some time $t$. Stream signals and background data
are encoded as facts, and temporal information 
by adding a time argument to atoms.
The central ideas of the encoding are illustrated by the following
example.
\begin{example}\label{ex:encoding-idea}
  Consider the LARS program $P$ comprising the single rule
  $r = b(X) \leftarrow \timeWindow{2} \Diamond a(X)$. Assume we are at
  time $t=7$. We 
  replace the window atom in the body by a fresh
  atom $\encWindowAtom(X)$, which must hold if $a(X)$ holds at $7$,
  $6$ or $5$. Thus, we can encode $r$ in ASP by the following rules:
  $b(X) \leftarrow \encWindowAtom(X); \encWindowAtom(X) \leftarrow
  \timePinned{a}(X,7); \encWindowAtom(X) \leftarrow
  \timePinned{a}(X,6); \encWindowAtom(X) \leftarrow
  \timePinned{a}(X,5)$. Assume an atom $a(y)$ was streaming in at
  time $5$; 
  modeled as time-pinned fact $\timePinned{a}(y,5)$, we derive
  $\encWindowAtom(y)$ and thus $b(y)$. That is, $b(y)$ holds at time
  $7$, since signal $a(y)$ at $5$ is still within the window.
\end{example}
Conceptually, the
translation of a LARS program $P$ to an
ASP program $\enc{P}$ is such that if atom $a(\vec{x})$ (where
$\vec{x}=x_1,\dots,x_n$) is in an answer set $A$ of $\enc{P}$, then
$a(x)$ holds \emph{now}. If the current time point is $t$, this is
encoded in two ways, viz.\ by $a(\vec{x}) \in A$ and
the time-pinned atom $\timePinned{a}(\vec{x},t) \in A$. This auxiliary atom
corresponds to the LARS $@$-atom
$@_t a(\vec{x})$, which then also holds now.
In general for any
$t' \in \bbN$, if $@_{t'} a(\vec{x})$ holds in an answer stream $S$ now,
then $\timePinned{a}(\vec{x},t')$ is 
in the corresponding
answer set $\enc{S}$, 
but  $a(\vec{x})$ is included only for $t'=t$.
The resulting equivalence is stated by the 
rules~$Q$
in
Alg.~\ref{alg:lars-program-to-asp}, Line~\ref{alg-line:lars-program-to-asp:Q}. To 
single out the current time point, we use an auxiliary predicate
$\now$. %

The ASP encoding $\enc{P}$ for $P$ at $t$ is then obtained by $Q$, %
$\{\now(t)\}$ and %
rule encodings $R$ as
computed by %
$\mi{larsToAspRules}$. Given a LARS rule $r$ of
form~\eqref{eq:rule-syntax}, we replace every non-ordinary extended atom
by a new auxiliary atom $\mi{atm}(e)$
(Lines~\ref{alg-line:atm:start}-\ref{alg-line:atm:end}).
Accordingly, for $e$ of form $@_T a(\vec{X})$, we use %
$\timePinned{a}(\vec{X},T)$ 
(where $T$ and $\vec{X}$ can be non-ground). For a window atom $e$, we
use a new predicate $\encWindowAtom_e$ for an \emph{encoded window atom}.
If $e$ has the form $\window^w\star\, a(\vec{X})$, $\star \in\{\Diamond,\Box\}$
, we use a new atom
$\encWindowAtom_e(\vec{X})$, while for 
$e$ of form $\window^w @_T a(\vec{X})$, we use
$\encWindowAtom_e(\vec{X},T)$ with a time argument. 

\leanparagraph{Window encoding}\label{par:window-encoding}
Predicate $\encWindowAtom_e$ has to hold in an
answer set $\enc{S}$ of $\enc{P}$ iff $e$ holds in a corresponding
answer stream $S$ of $P$ at $t$. We use 
the function $\mi{windowRules}$, which returns a
set of rules to derive $\encWindowAtom_e$
depending on the %
window (Lines~\ref{alg-line:rules:start}-\ref{alg-line:rules:end}).
In case $e = \timeWindow{n} @_T a(\vec{X})$ %
we %
have to test whether $\timePinned{a}(\vec{X},T)$
holds for some time $T$ within the last $n$ time points. For
$\timeWindow{n} \Diamond a(\vec{X})$, we omit $T$ in the rule
head. Dually, if $\timeWindow{n} \Box a(\vec{X})$ holds for the same
substitution $\vec{x}$ of $\vec{X}$ for all previous $n$ time points,
then in particular it holds now. So we derive
$\encWindowAtom_e(\vec{x})$ by the rule in
Line~\ref{alg-line:window-rules:time-box-static} if $a(\vec{x})$ holds
now and there is no {\em spoiler}\/ i.e., a time point among
$t-1,\dots,t-n$ where $a(\vec{x})$ does not hold. This is established by
the rule in Line~\ref{alg-line:window-rules:time-box-spoil}.
(We assume the window does not exceed the timeline and thus do not check
$T-i \geq 0$.)
Adding $a(\vec{X})$
to the body ensures safety of $\vec{X}$ in
$\timePinned{a}(\vec{X},T)$.
For $\tupleWindow{n} @_T a(\vec{X})$, we match every atom
$a(\vec{x})$ with
the time it occurs in the window of the last~$n$
tuples. Accordingly, we
track the relation between
arguments $\vec{x}$, the time $t$ of occurrence in the stream, and the
count $c$.
To this end, we assume any input signal $a(\vec{x})$ is provided as
$\{\timePinned{a}(\vec{x},t),\tickPinned{a}(\vec{x},t,c)\}$.
Furthermore, the rules in Line~\ref{alg-line:window-rules:tuple-at}
employ a predicate $\cnt$ that specifies the current tick count (as does
$\now$ for the time tick).  Based on this, the window is created
analogously to a time-based window but counting back $n-1$ tuples
instead of $n$ time points.
The case $\tupleWindow{n} \Diamond a(\vec{X})$ is again analogous, but
variable $T$ is not included in the head.

For $\tupleWindow{n} \Box a(\vec{X})$,
Line~\ref{alg-line:window-rules:tuple-box-static} is as in the
time-based analogue (Line~\ref{alg-line:window-rules:time-box-static});
$a(\vec{X})$ must hold now and there must not exist a spoiler. %
First, %
Line~\ref{alg-line:window-rules:tuple-box-spoil-time-range} ensures that
$a(\vec{X})$ holds at every time point~$T$ in the window's range, %
determined
by reaching back $n-1$ tick counts to count
$D$. %
To do so, we add
to the input stream an auxiliary atom of form $\tick(t,c)$ 
for every tick $(t,c)$ of the stream.
Second, %
Line~\ref{alg-line:window-rules:tuple-box-spoil-count-range} accounts
for the cut-off position within a time point, %
ensuring $a$ is within the selected range of counts.
Finally, ${\mi{windowRules}(e)=\emptyset}$ if $e$ is an atom or
an $@$-atom, as they do not need extra rules for their derivation.
\begin{example}\label{ex:tuple-box-static}
  Consider a stream $\tickStream'$, which adds to $\tickStream$ from
  Ex.~\ref{ex:tick-stream} tick $(4,3)$ with evaluation
  $\tickIntpr(4,3)=\{a\}$. We evaluate $\tupleWindow{2}\Box a$. The
  tick-pinned atoms are $\tickPinned{a}(3,1)$, $\tickPinned{b}(3,2)$ and
  $\tickPinned{a}(4,3)$; the window selects the last two, i.e., atoms
  with counts $D\geq 2$. It thus covers time points $3$ and $4$. While
  atom~$a$ occurs at time $3$, it is not included in the window anymore,
  since its count is $1<D$.
\end{example}

\leanparagraph{Stream encoding}
Let $O=(K,\tickIntpr)$ be a tick stream at tick $(t_m,c_m)$. We define
its encoding $\enc{O}$ as
$\{ \timePinned{a}(\vec{x},t) \mid {a(\vec{x}) \in \tickIntpr(t,c)}, {(t,c)
\in K} \} \cup
\{ \tickPinned{a}(\vec{x},t,c) \mid {a(\vec{x}) \in \tickIntpr(t,c)}, {(t,c) \in K}, {a(\vec{x}) \in \Atoms^\extensional}\} \cup
\{ \cnt(c_m)\} \cup
\{ \tick(t,c) \mid (t,c) \in K \}$.
We may assume that rules access background data $\backgroundData$
only by atoms (and not with $@$-atoms or window atoms).
Viewing $\backgroundData$ as facts in the program, we skip further
discussion.
The following %
implicitly disregards auxiliary atoms in the
encoding.
\begin{proposition}\label{prop:soundness-static-encoding}
  Let $P$ be a LARS program, $D=(K,\tickIntpr)$ be a tick data
  stream at tick $(t,c)$ and let
  $\enc{P}=\mi{LarsToAsp(P,t)}$. Then, $S$ is an answer stream of $P$
  for $D$ at $t$ iff $\enc{S}$ is an answer set of
  $\enc{P} \cup \enc{D}$.
\end{proposition}
\begin{example}\label{ex:asp-encoding-detail}
  We consider program $P$ of Example~\ref{ex:encoding-idea}, i.e., the
  rule $r=b(X) \leftarrow \timeWindow{2} \Diamond a(X)$. The translation
  $\enc{P}=\mi{LarsToAsp}(P,7)$ is given by the following rules, where
  $\encWindowAtom$ = $\encWindowAtom_{\timeWindow{2} \Diamond a(X)}$:
\begin{small}
$$
\begin{array}{lr@{~~}c@{~~}l}
     r_0: & b(X) & \leftarrow & \encWindowAtom(X)\\
     r_1: & \encWindowAtom(X) & \leftarrow & \now(\dot{N}),\,\timePinned{a}(X,T),\,T=\dot{N}-0\\
     r_2: & \encWindowAtom(X) & \leftarrow & \now(\dot{N}),\,\timePinned{a}(X,T),\,T=\dot{N}-1\\
     r_3: & \encWindowAtom(X) & \leftarrow & \now(\dot{N}),\,\timePinned{a}(X,T),\,T=\dot{N}-2\\
     r_n: &  now(7) & \leftarrow & 
\end{array}
\qquad
\begin{array}{lr@{~~}c@{~~}l}
     q_1: & a(X) & \leftarrow & \now(\dot{N}),\,\timePinned{a}(X,\dot{N})\\
     q_2: &  \timePinned{a}(X,\dot{N}) & \leftarrow & \now(\dot{N}),\,a(X)\\
     q_3: &  b(X) & \leftarrow & \now(\dot{N}),\,\timePinned{b}(X,\dot{N})\\
     q_4: &  \timePinned{b}(X,\dot{N}) & \leftarrow &
     \now(\dot{N}),\,b(X)\\    
        & &   
\end{array}
$$
\end{small}
The single answer stream of $P$ for $D$ at $7$ is
$I=([0,7],\{ 5 \mapsto \{a(y)\}, 7 \mapsto \{b(y)\}\}$) which
corresponds to the set
$\{\timePinned{a}(y,5),\timePinned{b}(y,7),b(y)\}$. In addition, the
answer set $\enc{S}$ of $\enc{P} \cup \enc{D}$ contains auxiliary
variables $\now(7)$, $\cnt(1)$,
$\tickPinned{a}(y,5,1)$
and $\omega(7)$
(and $\mi{tick}$ atoms).
\end{example}

\section{Incremental ASP Encoding}
\label{sec:incremental-encoding}
In this section, we present an incremental evaluation technique by
adjusting an incremental variant of the given ASP encoding. We illustrate
the central ideas in the following example.
\begin{example}[cont'd]\label{ex:incremental-encoding}
  Consider the following rules $\Pi$ similar to $\enc{P}$ of
  Ex.~\ref{ex:asp-encoding-detail} where predicate $\now$ is
  removed. Furthermore, we instantiate the \emph{tick time variable}
  $\dot{N}$ with $7$ to obtain so-called \emph{pinned} rules. (Later,
  pinning also includes grounding the \emph{tick count variable}
  $\dot{C}$ with the tick count.)
  \begin{small}
  \begin{displaymath}
    \begin{array}{lr@{~~}c@{~~}l@{\qquad\qquad}lr@{~~}c@{~~}l}
      r'_0: & b(X) & \leftarrow & \encWindowAtom(X)
      & q'_1: & a(X) & \leftarrow & \timePinned{a}(X,7) \\
      r'_1: & \encWindowAtom(X) & \leftarrow & \timePinned{a}(X,7)
      & q'_2: &  \timePinned{a}(X,7) & \leftarrow & a(X)\\
      r'_2: & \encWindowAtom(X) & \leftarrow & \timePinned{a}(X,6)
      & q'_3: &  b(X) & \leftarrow & \timePinned{b}(X,7)\\
     r'_3: & \encWindowAtom(X) & \leftarrow & \timePinned{a}(X,5)
     & q'_4: &  \timePinned{b}(X,7) & \leftarrow & b(X)
    \end{array}
  \end{displaymath}
  \end{small}  
  Based on the stream, encoded by
  $\enc{D}=\{\timePinned{a}(y,5), \tickPinned{a}(y,5,1)\}$ (we omit tick
  atoms), we obtain a ground program $\enc{P}_{D,(7,1)}$ from $\Pi$ by
  replacing $X$ with $y$; the answer set is
  $\enc{D} \cup \{ \encWindowAtom(y), b(y), \timePinned{b}(y,7)\}$.

  Assume now that time moves on to $t'=8$, i.e., a stream $D'$ at tick
  $(8,1)$. We observe that rules $q'_1,\dots,q'_4$ must be replaced by
  $q''_1,\dots,q''_4$, which replace time pin $7$ by $8$. Rule $r'_0$
  can be maintained since it does not contain values from ticks. The
  time window covers time points $6,7,8$. This is reflected by removing
  $r'_3$ and instead adding
  $\encWindowAtom(X) \leftarrow \timePinned{a}(X,8)$.

  That is, based on the time increment from $(7,1)$ to $(8,1)$, rules
  $E^- = \{q'_1,\dots,q'_4,r'_3\}$ and their groundings $G^-$ (with
  $X \mapsto y$) \emph{expire}, and new rules
  $E^+=\{q''_1,\dots,q''_4, \encWindowAtom(X) \leftarrow
  \timePinned{a}(X,8)\}$ have to be grounded based on the remaining
  rules (and the data stream), yielding new ground rules $G^+$.
  We thus incrementally obtain a ground program
  $\enc{P}_{D',(8,1)} = (\enc{P}_{D,(7,1)} \setminus G^-) \cup G^+$, which
  encodes the program $P$ for evaluation at tick $(8,1)$.
\end{example}
Before we formalize the illustrated incremental evaluation, we present
its ingredients.

\begin{algorithm}[t]
  \caption{Incremental Rules $\mi{IncrementalRules}(t,c,\signalSet)$}
  \label{alg:incremental-rules}
  \DontPrintSemicolon
  \SetAlgoVlined
  \small
  \KwIn{Tick time $t$, tick count $c$, signal set $\signalSet$ with at
    most one input signal, which is empty iff $(t,c)$ is a time
    increment. (The LARS program $P$ is global.)}
  \KwOut{Pinned incremental rules annotated with duration until expiration}

  \blockskip

      $F \assign \{\langle (\infty,\infty),\tick(t,c) \leftarrow \rangle \}$\;
    \lForEach{$a(\vec{x}) \in \signalSet\colon$}{$F \assign F \cup \{ \langle (\infty,\infty),\timePinned{a}(\vec{x},t) \leftarrow\rangle, \; \langle (\infty,\infty),\tickPinned{a}(\vec{x},t,c) \leftarrow \rangle \}$}
    {$Q \assign \{\, \langle (1,\infty),a(\vec{X}) \leftarrow
    \timePinned{a}(\vec{X},t) \rangle,\; \langle
    (1,\infty),\timePinned{a}(\vec{X},t) \leftarrow a(\vec{X}) \rangle \mid
    a \text{ is a predicate in }P\}$\;}
    $R \assign \emptyset$\;
    \ForEach{$r \in P$}{
      $\enc{r} \assign \mi{baseRule}(r)$\quad\algocomment{as defined in Alg.~\ref{alg:lars-program-to-asp}}\;
      $I \assign \bigcup_{e \in B(r)} \mi{incrementalWindowRules(e,t,c)}$\;
      $R \assign R \cup I \cup \{ \langle (\infty,\infty), \enc{r} \rangle \}$
    }

  \Return $F \cup Q \cup R$

  \blockskip
  \hrule
  \blockskip
  
  \DefnCustom{$\mi{incrementalWindowRules}(e,t,c)=\Match~e$\label{alg-line:incremental-window-rules:start}}{
      \Case~~$\timeWindow{n} @_{T} a(\vec{X}) \MapTo \{\, \langle (n+1,\infty), \encWindowAtom_e(\vec{X},t) \leftarrow \timePinned{a}(\vec{X},t) \rangle\,\}$\\
      \Case~~$\timeWindow{n} \Diamond a(\vec{X}) \MapTo \{\, \langle (n+1, \infty), \encWindowAtom_e(\vec{X}) \leftarrow \timePinned{a}(\vec{X},t) \rangle \,\}$\\
      \Case~~$\timeWindow{n} \Box a(\vec{X}) \MapTo\{\, \langle (\infty,\infty), \encWindowAtom_e(\vec{X}) \leftarrow a(\vec{X}),\naf\, \mi{spoil}_e(\vec{X}) \rangle\,\}\,\cup$\\
      \quad $\{\, \langle (n,\infty), \mi{spoil}_e(\vec{X}) \leftarrow a(\vec{X}), \naf\, \timePinned{a}(\vec{X},t-1) \rangle \,\}$\label{alg-line:incr-time-box}\quad\algocomment{{\small only if $n \geq 1$}}\\
      %
      %
      %
      %
      \Case~~$\tupleWindow{n} @_{T} a(\vec{X}) \MapTo \{\, \langle (\infty,n),\encWindowAtom_e(\vec{X},t) \leftarrow \tickPinned{a}(\vec{X},t,c) \rangle\,\}$\\
      \Case~~$\tupleWindow{n} \Diamond a(\vec{X}) \MapTo \{\, \langle (\infty,n),\encWindowAtom_e(\vec{X}) \leftarrow \tickPinned{a}(\vec{X},t,c) \rangle \,\}$\\
      \Case~~$\tupleWindow{n} \Box a(\vec{X}) \MapTo\{\, \langle (\infty,\infty),\encWindowAtom_e(\vec{X}) \leftarrow a(\vec{X}),\naf\, \mi{spoil}_e(\vec{X})\rangle\,\}\,\cup$\\
      \quad $\{\, \langle (\infty,n),\mi{spoil}_e(\vec{X}) \leftarrow a(\vec{X}), \tick(t,c), \mi{covers}^\wfnTime_e(t),\, \naf\, \timePinned{a}(\vec{X},t) \rangle\,\}\,\cup$\label{alg-line:incr-tuple-box-spoil-covers-time}\\
      \quad $\{\, \langle (\infty,n),\mi{spoil}_e(\vec{X}) \leftarrow \tickPinned{a}(\vec{X},t,c), \mi{covers}^\wfnTime_e(t), \,\naf\,\mi{covers}^\wfnTuple_e(c) \rangle\,\}\,\cup$\label{alg-line:incr-tuple-box-spoil-covers-count}\\
      \quad $\{\, \langle (\infty,n), \mi{covers}^\wfnTime_e(t) \leftarrow \tick(t,c) \rangle \;, \langle (\infty,n), \mi{covers}^\wfnTuple_e(c) \leftarrow \tick(t,c) \rangle \,\}$\label{alg-line:incr-tulpe-box-covers}\\
      \lElse{$~\emptyset$}\label{alg-line:incremental-window-rules:end}
  }
\end{algorithm}

\leanparagraph{Algorithm~\ref{alg:incremental-rules}: Incremental rule
  generation}
Alg.~\ref{alg:incremental-rules} shows the procedure
$\mi{IncrementalRules}$ that obtains incremental rules based on a tick
time $t$, a tick count $c$, and the \emph{signal set}
$\signalSet = \tickIntpr(t,c)$, where $\signalSet=\emptyset$, if
$(t,c)$ is a time increment of $k$. The resulting rules of
Alg.~\ref{alg:incremental-rules} are annotated with a tick that
indicates how long the ground instances of these rules are applicable
before they expire.
\begin{definition}[Annotated rule]
  Let $(t,c)$ be a tick, where $t,c \in \bbN \cup \{\infty\}$, and $r$
  be a rule. Then, the pair $\langle (t,c), r \rangle$ is called an
  \emph{annotated rule}, and $(t,c)$ the \emph{annotation} of $r$.
\end{definition}
Annotations serve two purposes. First, in
Alg.~\ref{alg:incremental-rules}, they express a \emph{duration} how
long a generated rule is applicable. Then, in
Alg.~\ref{alg:increment-tick} below this duration will be added to the
current tick to obtain the \emph{expiration tick} (annotation) of a
rule. If a rule \emph{expires} at tick $(t,c)$, i.e., if 
its expiration tick $(t',c')$ fulfills $t'\geq t$ or $c' \geq c$, then it
has to be deleted from the encoding.
\begin{example}[cont'd]\label{ex:annotations}
  Each rule $q'_i$, $1 \leq i \leq 4$, has duration $(1,\infty)$. That
  is, after 1 time point, the rule will expire, regardless of how many
  atoms appear at the current time point. Hence, the \emph{time
    duration} is $1$, and the \emph{count duration} is infinite, since
  these rules cannot expire based on arrival of atoms. Similarly, rules
  $r'_i$, $1 \leq i \leq 3$, have duration $(2,\infty)$ due to the time
  window length $2$.
\end{example}
We will discuss expiration ticks based on these durations
below. Algorithm~\ref{alg:incremental-rules} is concerned with
generating the incremental rules and their durations.
In the first two lines, auxiliary facts, as discussed earlier, are added
to a fresh set $F$. These facts expire neither based on time nor count,
hence the duration annotation $(\infty,\infty)$. As illustrated in
Ex.~\ref{ex:annotations}, we collect in set $Q$ the incremental analogue
of $Q$ in Alg.~\ref{alg:lars-program-to-asp}. These rules expire after 1
time point, hence the annotation $(1,\infty)$.

Within the loop we collect for every LARS rule $r$ a base rule $\enc{r}$
(as in Alg.~\ref{alg:lars-program-to-asp}), together with incremental
window rules, computed 
by $\mi{incrementalWindowRules}$
(Lines~\ref{alg-line:incremental-window-rules:start}-\ref{alg-line:incremental-window-rules:end}). We
assign an infinite duration $(\infty,\infty)$ to the base rule $\enc{r}$
since it never needs to expire, i.e., it suffices to ensure that encoded
window atoms $\encWindowAtom_e$ expire correctly. An optimized version
may expire also 
$\enc{r}$ due to the durations of atoms $\encWindowAtom_e$
from the incremental windows that derive them.

\leanparagraph{Incremental window encoding}
We already gave the intuition for 
atoms $\timeWindow{n} \Diamond a(\vec{X})$. The case of
$\timeWindow{n} @_T a(\vec{X})$ is similar. Like in the static
translation, we additionally have to use the time information in the
head. Similarly, $\tupleWindow{n} \Diamond a(\vec{X})$ and
$\tupleWindow{n} @_T a(\vec{X})$ expire after $n$ new incoming atoms,
instead of $n$ time points.
For $\timeWindow{n} \Box a(\vec{X})$, we add a spoiler rule for the previous
time point $t-1$, which will be considered for the next $n$ time
points.

For $e=\tupleWindow{n} \Box a(\vec{X})$ we maintain two spoiler rules as
in the static case
that ensure $a(\vec{X})$ occurs at all time
points in the coverage of the window, and the occurrence of
$a(\vec{X})$ at the leftmost time point is also covered 
by the tick count.
At tick $(t,c)$, we have a guarantee for the next~$n$ atoms that tick
time $t$ will be covered within the window. This is expressed by a rule
$\mi{covers}^\tau_e(t) \leftarrow \tick(t,c)$ with duration
$(\infty,n)$. Likewise, $\mi{covers}^\#_e(c) \leftarrow \tick(t,c)$ will
select tick count $c$ within duration $(\infty,n)$. Notably, coverage
for time increments $(t+k,c)$ may extend the tuple window arbitrarily long if
no atoms appear. As the spoiler rules are based on these cover
atoms, their expiration is optional, i.e., keeping them does not yield
incorrect inferences. However, we can also expire them when they become
redundant, i.e., after $n$ atoms.
Finally, $\mi{IncrementalRules}$ returns the $F \cup Q \cup R$, where $R$
contains all base rules and incremental window rules.

\begin{algorithm}[t]
  \caption{Single tick increment $\mi{IncrementTick}(\Pi,G,t,c,\signalSet)$}
  \label{alg:increment-tick}
  \DontPrintSemicolon
  \SetAlgoVlined
  \small
  \KwIn{Set of annotated, cumulative incremental rules
    $\Pi \supseteq \enc{D}$ collected until previous tick; its annotated
    groundings
    $G = \bigcup_{\langle (t',c'),r \rangle \in \Pi}
    \mi{ground}(\Pi,r)$, tick time $t$, tick count~$c$ and signal set
    $\signalSet$}
  \KwResult{Updated $\Pi$ and $G$}

  \blockskip
  
  $I \assign \mi{IncrementalRules}(t,c,\signalSet)$\;
  $E^+ \assign \{\langle (t+t_\Delta, c+c_\Delta), r \rangle \mid \langle (t_\Delta,c_\Delta), r \rangle \in I\}$\quad\algocomment{\footnotesize determine expiration for new rules}\;\label{alg-line:increment-tick:assign-E-plus}
  $E^- \assign \{ \langle (t',c'),r \rangle \in \Pi \mid t' \leq t \text{ or }c' \leq c\}$\quad\algocomment{\footnotesize expired incremental rules}\;\label{alg-line:increment-tick:assign-E-minus}
  $\Pi' \assign (\Pi \setminus E^-) \cup E^+$\;\label{alg-line:increment-tick:update-Pi}
  $G^+ \assign \{ \langle (t',c'), r' \rangle \mid \langle (t',c'), r \rangle \in E^+, r' \in \mi{ground}(\Pi',r)\}$\quad\algocomment{\footnotesize \footnotesize new ground rules with expiration}\;\label{alg-line:increment-tick:assign-G-plus}
  $G^- \assign \{ \langle (t',c'),r \rangle \in G \mid t' \leq t \text{ or } c' \leq c\}$\quad\algocomment{\footnotesize expired ground rules with expiration annotation}\;\label{alg-line:increment-tick:assign-G-minus}
  $G' \assign (G \setminus G^-) \cup G^+$\;\label{alg-line:increment-tick:update-G}  
  \Return $\langle \Pi', G' \rangle$
\end{algorithm}

\leanparagraph{Algorithm~\ref{alg:increment-tick}: Incremental evaluation}
Alg.~\ref{alg:increment-tick} gives the high-level procedure
$\mi{IncrementTick}$ to incrementally adjust a program encoding.
We assume the function $\mi{ground}(\Pi,r)$ returns all possible ground
instances of a rule $r\in \Pi$ 
(due to constants in~$\Pi$). In fact, $\mi{IncrementTick}$ maintains a program
$\Pi$ that contains the encoded data stream~$\enc{D}$ and non-expired
incremental rules as obtained by consecutive calls to
$\mi{IncrementalRules}$, tick by tick. Moreover,
it maintains a grounding~$G$ of~$\Pi$, i.e., the incremental encoding for
the previous tick plus expiration annotations.

The procedure starts by generating the new incremental rules~$I$ based
on Alg.~\ref{alg:incremental-rules} described above. Next, we add for
each rule the current tick~$(t,c)$ to its duration~$(t_\Delta,c_\Delta)$
(componentwise). This way, we obtain new incremental rules~$E^+$ with
expiration tick annotations. Dually, we collect in~$E^-$ previous
incremental rules that expire now, i.e., when the current
tick reaches the expiration tick
time~$t'$ or 
count~$c'$. The new cumulative program $\Pi$ 
results by removing $E^-$
from $\Pi$ and 
adding $E^+$. Based on
$\Pi'$,
we obtain in Line~\ref{alg-line:increment-tick:assign-G-plus} the new
(annotated) ground rules~$G^+$ based on~$E^+$. As in
Line~\ref{alg-line:increment-tick:assign-E-minus}, we determine in
Line~\ref{alg-line:increment-tick:assign-G-minus} the set~$G^-$ of
expired (annotated) ground rules. After assigning $G'$  the updated annotated
grounding in Line~\ref{alg-line:increment-tick:update-G}, we
return the new incremental evaluation state $\langle \Pi',G' \rangle$,
from which the current incremental program is derived as follows.
\begin{definition}[Incremental Program]
  Let $P$ be a LARS program and $D=(K,\tickIntpr)$ be a tick stream,
  where $K=\langle (t_1,c_1),\dots,(t_m,c_m)\rangle$. The \emph{incremental program $\enc{P}_{D,k}$ of $P$ for $D$ at tick $(t_k,c_k)$}, $1 \leq k \leq m$, is defined by  $\enc{P}_{D,k}=\{r \mid \langle (t',c'), r \rangle \in G_k\}$, where
  \begin{displaymath}
    \langle \Pi_k, G_k \rangle = 
    \begin{cases}
      \mi{IncrementTick}(\emptyset,\emptyset,t_1,c_1,\emptyset) & \text{if}~k=1,\\
      \mi{IncrementTick}(\Pi_{k-1},G_{k-1},t_k,c_k,\tickIntpr(t_k,c_k)) & \text{else.}
    \end{cases}
  \end{displaymath}
\end{definition}
In the following, body occurrences of form $@_t a(\vec{X})$ are viewed as shortcuts for $\window^\infty @_t a(\vec{X})$.
The next proposition states that to faithfully compute an incremental
program from scratch, it suffices to start iterating
$\mi{IncrementalTick}$ from the oldest tick that is covered from any
window in the considered program.
In the subsequent results we disregard auxiliary atoms like
$tick(t,c), \mi{covers}^\tau_e(t)$, etc.
Let ${\AS^\intensional(\enc{P})}$ denote
the answer sets of $\enc{P}$, projected to intensional atoms.
\begin{proposition}\label{prop:cut-off}
  Let $D=(K,\tickIntpr)$ and $D'=(K',\tickIntpr')$ be two data streams such that
  (i) $D' \subseteq D$, (ii)
  $K=\langle (t_1,c_1), \dots, (t_m,c_m)\rangle$ and (iii)
  $K'=\langle(t_k,c_k), \dots, (t_m,c_m)\rangle$, $1 \leq k \leq
  m$. Moreover, let $P$ be a LARS program and $n^\tau$ (resp.\ $n^\#$) be
  the maximal window length for all time (resp.\ tuple) windows;
  or $\infty$ if none exists.
  If $t_k \leq t_m-n^\tau$ and $c_k \leq c_m-n^\#+1$, then
  $\AS^\intensional(\enc{P}_{D,m}) =
  \AS^\intensional(\enc{P}_{D',m})$.
\end{proposition}
The result stems from the fact that in the incremental program
$\enc{P}_{D,m}$ no rule can fire based on outdated information, i.e.,
atoms that are not covered by any window anymore. In order to obtain an
equivalence between $\enc{P}_{D,m}$ and $\enc{P}_{D',m}$ on extensional
atoms, we would have to drop all atoms of the stream encoding $\enc{D}$
during $\mi{IncrementalTick}$, as soon as no window can access them
anymore.

The following states the correspondence between the static and 
the incremental encoding.
\begin{proposition}%
  \label{prop:equiv-static-and-incremental-encoding}
  Let $P$ be a LARS program and $D$ be a tick data stream
  at tick $m=(t,c)$. Furthermore, let $\enc{P}=\mi{LarsToAsp(P,t)}$ and
  $\enc{P}_{D,m}$ be the incremental program at tick
  $m$. Then $S \cup \{\now(t),\cnt(c)\}$ is an answer set of
  $\enc{P} \cup \enc{D}$ iff $S$ is an answer set of $\enc{P}_{D,m}$ (modulo aux. atoms).
\end{proposition}
In conclusion, we obtain from
Props.~\ref{prop:soundness-static-encoding} and \ref{prop:equiv-static-and-incremental-encoding}
the desired correctness of the incremental encoding.
\begin{theorem}%
  \label{theorem:incremental-lars-evaluation}
  Let $P$ be a LARS program and $D=(K,\tickIntpr)$ be a tick data stream
  at tick $m=(t,c)$. Then, $S$ is an answer stream of $P$ for $D$ at $t$ iff
  $\enc{S}$ is an answer set of $\enc{P}_{D,m}$ (modulo aux. atoms).
\end{theorem}

\section{Implementation}\label{sec:implementation}
We now present \emph{Ticker}, our stream reasoning engine which is
written in \emph{Scala} (source code available at
\url{https://github.com/hbeck/ticker}).
It has two high-level processing methods for a given time point:
\emph{append} is adding input signals, and \emph{evaluate} returns the
model.
Two implementations of this interface are provided, based on two
evaluation strategies discussed next.

\leanparagraph{One-shot solving by using Clingo}\label{sec:implemenation:clingo}
The ASP solver Clingo~\cite{clingo14} is a practical choice for
stratified programs, where no ambiguity arises which model to compute. %
At every time point, resp., at the arrival of a new atom,
the static LARS encoding $\enc{P}$ (of
Alg.~\ref{alg:lars-program-to-asp}) is streamed to the solver and
results are parsed as soon as Clingo reports a model. In case of
multiple models, we take the first one. Apart from this so-called
push-based mode, where a model is prepared after every \emph{append}
call, we also provide a pull-based mode, where only \emph{evaluate}
triggers model computation.
\ifextended
As argued in \ref{sec:reactive-clingo}, Clingo's reactive features are not
applicable.
\fi
 
\leanparagraph{Incremental evaluation by
  TMS}\label{sec:implemenation:incremental}
In this strategy, the model is maintained continuously using our own
implementation of the truth-maintenance system (TMS) by~\cite{Doyle79}.
A TMS \emph{network} can be seen as logic program $P$ and data
structures that reflect a so-called \emph{admissible} model~$M$ for
$P$. Given a rule $r$, the network is updated such that it represents an
admissible model $M'$ for $P \cup \{r\}$, thereby reconsidering the
truth value of atoms in $M$ only if they may change due to the
network. Ticker analogously allows for rule removals, i.e., obtaining an
admissible model $M'$ for $P \setminus \{r\}$. We exploit the following
correspondence of admissible models and answer sets.
\begin{theorem}[cf. \cite{elkan90}]\label{thm:tms-asp-correspondence-main}
  (i) A model ${M}$ is admissible for program $P$ iff it is an
  answer set of~$P$. (ii) Deciding whether $P$ has an
  admissible %
  model is NP-complete.
\end{theorem}
Notably, this correspondence holds only in the
absence of constraints; or more generally, odd loops~\cite{elkan90}.
In case such programs are used, neither a correct output nor termination
are guaranteed.
Elkan points out that also incremental reasoning is NP-complete,
  i.e., given an admissible %
  model~$M$ for~$P$, deciding for a rule~$r$ whether ${P \cup \{r\}}$
  has an admissible %
  model.
  No further knowledge about TMS is required for our purpose. A
  detailed, formal review can be found in~\cite{jtms-rr17},
  supplementing the textual presentation in~\cite{Doyle79}.

When new data is streaming in, we compute the incremental rules $G^+$ as
defined in Alg.~\ref{alg:incremental-rules}, add them to the TMS
network, and remove expired ones $G^-$; which results in an immediate
model update.
The incremental TMS strategy is, due to its maintenance outset, more
amenable to keep the latest model by inertia, which may be desirable in
some applications.

\leanparagraph{Pre-grounding}
In Alg.~\ref{alg:increment-tick}, we assume a grounder that instantiates
pinned rules from Alg.~\ref{alg:incremental-rules}. To provide
according efficient techniques is a topic on its own;
we restrict grounding to the pinning process in
Alg.~\ref{alg:incremental-rules}.
To this end, we add to each rule for every variable $X$ in the scope of
a window atom an additional \emph{guard} atom that includes $X$. The
guard is either background data or intensional. Based on this, the
incremental rules in Alg.~\ref{alg:incremental-rules} can be grounded
upfront, apart from the tick variables~$\dot{N}$ and~$\dot{C}$ and time
variables in $@$-atoms. We call such programs \emph{pre-grounded}.
A LARS program $P$ is first translated into an encoding $\enc{P}$ with
several data structures that differentiate~$Q$, base rules~$R$, and
window rules~$W$. During the initialization process, pre-groundings are
prepared, where arithmetic expressions are represented by auxiliary
atoms. During grounding, they are removed if they hold, otherwise the
entire ground rule is removed.
\begin{example}\label{ex:pre-grounding}
  For rule
  $r=@_T \alphahigh \leftarrow \mi{value}(V), \timeWindow{n} @_T
  \mi{alpha}(V), V\geq 18$ of Ex.~\ref{ex:alpha-high-abstraction}, where
  $\mi{value}(V)$ was added as guard, we get a base rule
  $\enc{r}=\timePinned{\alphahigh}(T) \leftarrow \mi{value}(V),
  \omega_e(V,T), \mi{Geq}(V,18)$, where $e=\timeWindow{n}@_T\mi{alpha}(V)$.
  Given facts $\{\mi{value}(0),\dots,\mi{value}(30)\}$ (from background
  data or potential derivations), we obtain the pre-grounding
  $\{\timePinned{\alphahigh}(T) \leftarrow \mi{value}(x), \omega_e(x,T)
  \mid x \in \{18,\dots,30\}\}$.~~ %
\end{example}
We then use pre-groundings in Alg.~\ref{alg:incremental-rules} such that
when Alg.~\ref{alg:increment-tick} receives its result $I$, all rules
are already ground. Thus, the implementation has no further grounding in
Alg.~\ref{alg:increment-tick} and only concerns handling durations and
expirations, which is realized based on efficient lookups.

\section{Evaluation}\label{sec:evaluation}

For an experimental evaluation, we consider two scenarios in the context
of content-centric network management, where smart routers need to
manage packages dynamically~\cite{ccn17}.

\leanparagraph{Scenario A: Caching Strategy}
Fig.~\ref{fig:program:caching-strategy} shows a program to dynamically
select one of several strategies (\mi{fifo}, \mi{lfu}, \mi{lru},
\mi{random}) how to replace content items (video chunks) in a local
cache. A user request parameter~$\alpha$, signaled as atom
$\alphaval(V)$, is monitored and abstracted to a qualitative level
($r_1$-$r_3$) using tuple-based windows. At this level, time-based
windows are used to decide among \mi{fifo}, \mi{lfu}, and \mi{lru}
($r_4$-$r_6$); the default policy is \mi{random} ($r_7$-$r_{10}$).

Setup \emph{A1} replaces tuple windows in rules $r_1$--$r_3$ by time
windows (as in~\cite{ccn17}), setup \emph{A2} uses the program as
shown.
The input signals $\mi{alpha}(V)$ are generated such that a random mode
\emph{high}, \emph{medium} or \emph{low} is repeatedly chosen and kept
for twice the window size.

\leanparagraph{Scenario B: Content Retrieval}
Fig.~\ref{fig:program:retrieval} depicts the second program, which, in
contrast to the former, may have multiple models and includes
recursive computation, instead of straightforward chaining.
In a network, items can be cached and requested at every node. If a user
recently requested item $I$ at node $N$ (rule $r_1$), it is either
available at $N$ ($r_2$) or has to be retrieved from some other node $M$
($r_3,r_6$). A single node is selected ($r_3$) that provides the best
quality level (e.g.\ connection speed) among all reachable nodes having
$I$ ($r_5$). Connecting paths ($r_7$, $r_8$) work unless the end node of
an edge was down during the last~$n$ time points ($r_9$).
Finally, nodes repeatedly report their quality level, among which the
best recent value is selected ($r_{10}$).
We take the classic Abilene network~\cite{Spring2004}, i.e., the set of
edges $\{(x,y), (y,x) \mid (x,y) \in E\}$, where
$E=\{ (0,1),(1,2),\dots,(9,10),(0,10),(1,10),(2,8),(3,7)\}$. We use
three quality levels $\{0,1,2\}$ and two items.
In setup \emph{B1}, at every time point, with respective probability
$p=0.1$, each item is requested at a random node, one random item is
cached at a random node, and one random node is signalled as
down. Further, the quality level of each node changes with $p=3/n$,
where $n$ is the window size. Setup \emph{B2} requests each item with
$p=0.5$ at 1-3 random nodes, always signals 1-3 random cache entries,
and a quality level for every node with $p=0.25$, which is then with
$p=0.9$ the previous one. With $p=1/n$, a random node will be down for
$1.5 \cdot n$ time points.

\leanparagraph{Evaluations}
For each scenario and setup, we ran two evaluation modes. The first one
fixes the number $\mi{tp}$ of time points and increases the window size
$n$ stepwise; the second setup vice versa.

In each evaluation mode, we measure
\begin{inparaenum}[(i)]
\item the time $\tInit$ needed to initialize the engine before input
  signals are streamed (in case of the incremental mode, this includes
  pre-grounding),
\item the average time $\tTick$ per tick, i.e., a time or count
  increment, and
\item the total time $\tTotal$ of a single run, resulting from
  $\tInit$ and $\tTick$ for all timepoints and atoms.
\end{inparaenum}
(Note that a tick increment may involve both adding and removing
rules.)
Each evaluation includes runtimes for both reasoning strategies, i.e.,
based on Clingo (Vers. 5.1.0) and based on the incremental approach with
Doyle's TMS.
For a fair comparison with TMS, we use Clingo in a push-based
mode, i.e., a model is computed whenever a signal streams in.
To obtain robust results, we first run each instance twice without
recording time, and then build the average over the next 5 runs for
$\tInit$, $\tTotal$ and $\tTick$, respectively. The first two runs
serve as warm-up for the environment, ensuring that potential optimizations by the Java-Virtual-Machine (JVM)
do not distort the measurements.
All evaluations were executed on a laptop with an Intel i7 CPU at
2.7~GHz and 16~GB RAM running the JVM version
1.8.0\_112. They can be run via class {\small
  \texttt{LarsEvaluation}}. 

\begin{figure*}[t]
	\centering
  \begin{displaymath}
  \begin{array}{lr@{~~}c@{~~}llr@{~~}c@{~~}l}
	 r_1: & @_T\, \alphahigh  & \leftarrow & \mi{value}(V),\, \tupleWindow{n} @_T\, \alphaval(V),~ {18 \leq V} &
  	r_6: & \fifo & \leftarrow & \timeWindow{n}\Box\, \alphalow\\
	 r_2: & @_T\, \alphamid  & \leftarrow & \mi{value}(V),\, \tupleWindow{n} @_T\, \alphaval(V),~ {12 \leq V < 18} &
	  r_7: & \done & \leftarrow & \lfu \\
   r_3: & @_T\, \alphalow & \leftarrow & \mi{value}(V), \tupleWindow{n} @_T\, \alphaval(V),~ {V < 12} &
    r_8: & \done & \leftarrow & \lru \\
   r_4: & \lfu  & \leftarrow & \timeWindow{n} \Box\, \alphahigh &
    r_9: & \done & \leftarrow & \fifo \\
   r_5:& \lru & \leftarrow & \timeWindow{n} \Box\, \alphamid &
    r_{10}:&  \random & \leftarrow & \naf\, \done
  \end{array}
  \end{displaymath}	
	\vspace{-2ex}
	\caption{Program for Scenario A, Setup \emph{A2}. Setup \emph{A1} uses
    $\timeWindow{n}$ in $r_1-r_3$ instead of $\tupleWindow{n}$.}
	\label{fig:program:caching-strategy}
\end{figure*}

\begin{figure*}[t]
	\centering
  \begin{displaymath}
  \begin{array}{lr@{~~}c@{~~}l}
    r_1: & \need(I,N) & \leftarrow & \itemm(I),\, \node(N),\, \timeWindow{n} \Diamond  \req(I,N)\\
    r_2: & \avail(I,N) & \leftarrow & \itemm(I),\, \node(N),\, \timeWindow{n} \Diamond \cache(I,N)\\
    r_3: & \getFrom(I,N,M) & \leftarrow & \source(I,N,M),\, \naf\, \nGetFrom(I,N,M)\\
    r_4: & \nGetFrom(I,N,M) & \leftarrow & \node(M),\,\getFrom(I,N,M'),\,M \neq M'\\
    r_5: & \nGetFrom(I,N,M) & \leftarrow & \source(I,N,M),\,\source(I,N,M'),\,M \neq M',\,\qual(M,L),\,\qual(M',L'),\,L<L'\\
    r_6: & \source(I,N,M) & \leftarrow &  \need(I,N),\, \naf\, \avail(I,N),\, \avail(I,M),\,\reach(N,M)\\
    r_7: & \reach(N,M) & \leftarrow & \conn(N,M)\\
    r_8: & \reach(N,M) & \leftarrow & \reach(N,M'),\, \conn(M',M),\, M' \neq M,\, N \neq M\\
    r_9: & \conn(N,M) & \leftarrow & \edge(N,M),\, \naf\; \timeWindow{n} \Box\, \down(M)\\
    r_{10}: & \qual(N,L) & \leftarrow &  \node(N),\,\level(L),\,\level(L'),\,L' < L,\,\timeWindow{n} \Diamond \qLev(N,L),\,\naf\, \timeWindow{n} \Diamond \qLev(N,L')
  \end{array}
  \end{displaymath}
	\vspace{-2ex}
	\caption{Program for Scenario B}
	\label{fig:program:retrieval}
\end{figure*}

\leanparagraph{Results}
We report here on findings regarding the total execution times
$\tTotal$, shown in
Figures~\mbox{\ref{fig:results:caching:winsize:total}-\ref{fig:results:content:timepoints:total}}.
\ifextended
Detailed runtimes for $\tTotal$, $\tInit$ and $\tTick$ can be found in
Tables~\ref{tab:tp:caching:A1}--\ref{tab:winsize:content:B2} in the
Appendix.
\fi

Figures~\ref{fig:results:caching:winsize:total}-\ref{fig:results:content:winsize:total}
show the effect on the runtime when the window size is increased. We
observe that for both scenarios the total execution time $\tTotal$ is
proportionally growing using Clingo, while for the incremental
implementation (TMS) $\tTotal$ remains nearly constant. For Clingo, this
is explained by the full recomputation of the model with all previous
input data, while TMS benefits from prior model computations and is thus
significantly faster for larger window sizes.
Dually, Figures~\ref{fig:results:caching:timepoints:total}-\ref{fig:results:content:timepoints:total} show the runtime evaluation
for increasing number of timepoints. For both scenarios the total run
time $\tTotal$ of both Clingo and TMS increases linearly, and
incremental is significantly faster than repeated one-shot solving.
For both evaluations, using different windows (\emph{A1}
vs. \emph{A2}) has no influence on the execution time, for both Clingo
and TMS, and different input patterns (\emph{B1} vs. \emph{B2}) seem to
influence TMS less than Clingo.

In conclusion, the experiments indicate that incremental model update
may computationally pay off in comparison to repeated recomputing from
scratch, in particular when using large windows. Furthermore,
maintenance aims at keeping a model by inertia, which however we have
not assessed in the experiments.

\newcommand{\textwidthFactor}{0.67}

\begin{figure}[t!]
  \centering
  \includegraphics[width=\textwidthFactor\textwidth]{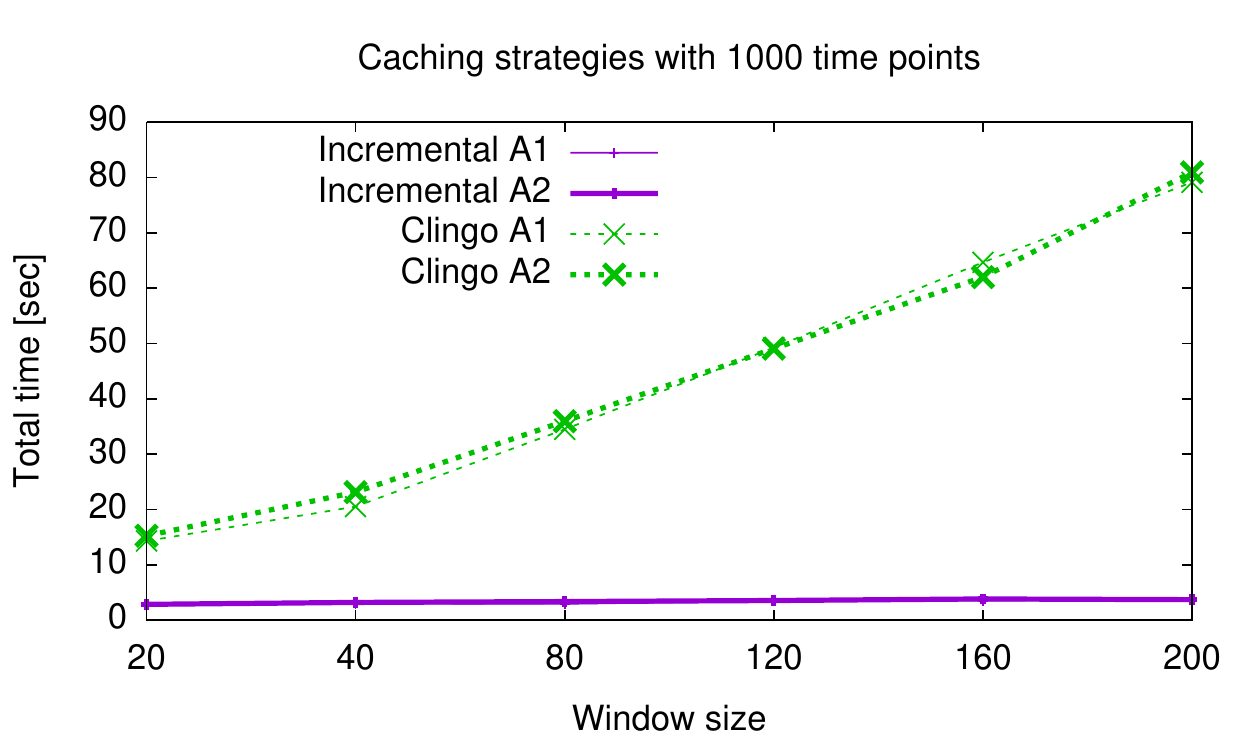}
  \caption{Runtime evaluation for increasing window size: Scenario A (Caching Strategy)} \label{fig:results:caching:winsize:total}
\end{figure}

\begin{figure}[t!]
  \centering
  \includegraphics[width=\textwidthFactor\textwidth]{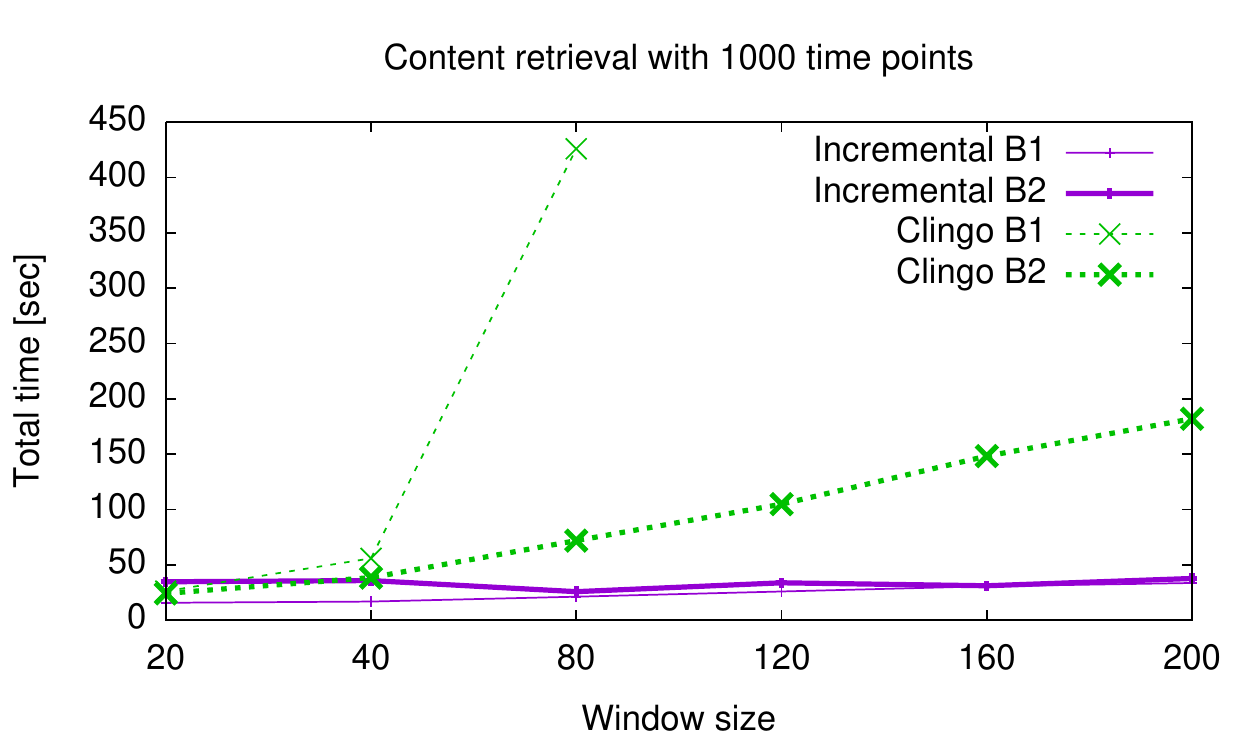}
  \caption{Runtime evaluation for increasing window size: Scenario B (Content Retrieval)}
  \label{fig:results:content:winsize:total}
  %
\end{figure}

%

\begin{figure}[t!]
  \centering
  \includegraphics[width=\textwidthFactor\textwidth]{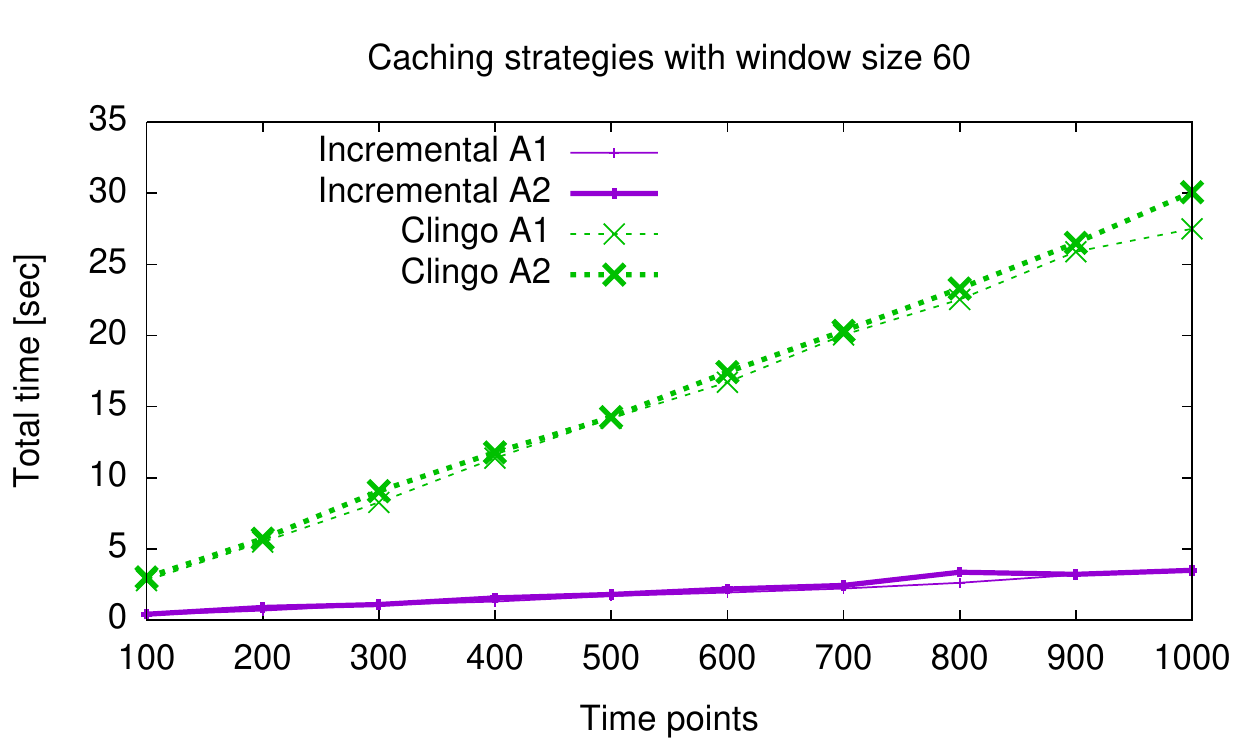}
  \caption{Runtime evaluation for increasing timepoints: Scenario A (Caching Strategy)}
  \label{fig:results:caching:timepoints:total}
\end{figure}

\begin{figure}[t!]  
  \centering
  \includegraphics[width=\textwidthFactor\textwidth]{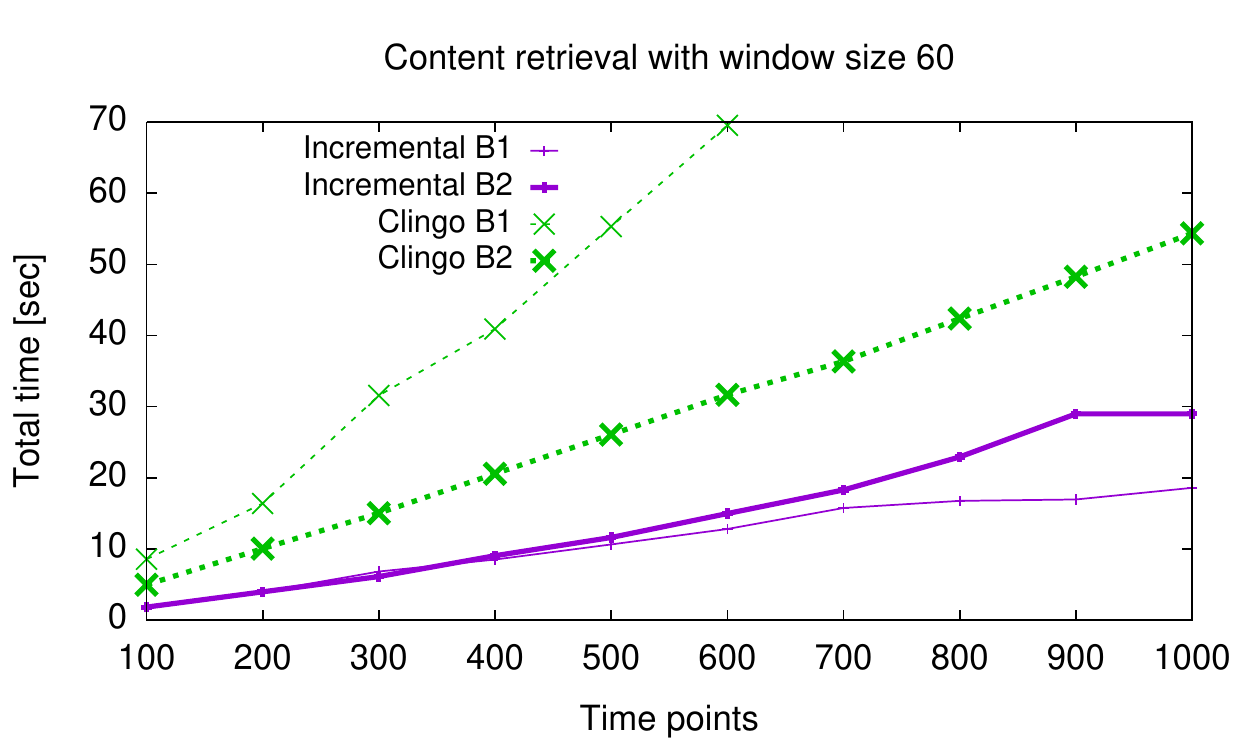}
  \caption{Runtime evaluation for increasing timepoints: Scenario B (Content Retrieval)}
  \label{fig:results:content:timepoints:total}
  %
\end{figure}


\section{Related Work and Conclusion}
\label{sec:related-work}

In~\cite{bde2015-ijcai}, TMS techniques have been extended
and applied for (plain) LARS, instead of reducing LARS to ASP. In
contrast, the present approach does not primarily focus on model update,
but incremental program update.
Apart from work on Clingo mentioned earlier, alternatives to one-shot
ASP were also considered by
\citeN{AlvianoDR14}.
The ASP approach of~\citeN{DoLL11} for stream reasoning calls the dlvhex
solver; it has no incremental reasoning and cannot handle heavy data
load.
ETALIS~\cite{arfs2012} is a prominent rule formalism for complex event
processing %
to reason about intervals for atomic events with a peculiar minimal model
semantics.
ETALIS is monotonic for a growing timeline (as such trivially incremental),
 and does not feature window mechanisms.
StreamLog \cite{Zaniolo12} extends Datalog for single-model stream
reasoning, where rules concluding about the past are excluded;
neither windows nor incremental evaluation were considered.
The DRed algorithm \cite{GuptaMS93Mod} for incremental Datalog update
deletes all consequences of deleted facts and then adds all rederivable
ones from the rest. It was adapted to RDF streams by
\citeN{Barbieri10}, where tuples are tagged with an expiration
time. 
\citeN{RenP11} explored TMS techniques for 
ontology streams. However, windows and time reference were not
considered in their monotonic setting.
Towards incremental grounding, techniques as
in~\cite{asperix,lazy-grounding-iclp09,omiga} might be considered.

\leanparagraph{Outlook}
The algorithms we have presented center around the idea of incrementally
adapting a model based on an incremental adjustment of a program.
Our implementation indicates performance benefits arising from
incremental evaluation. Developing techniques for full grounding
on-the-fly in this context remains to be done. On the semantic side,
notions of closeness between consecutive models and guarantees to obtain
them are intriguing issues for future work.

\leanparagraph{Acknowledgements}
We thank Roland Kaminski for providing guidance on the use of Clingo.

\bibliographystyle{acmtrans}

\ifinlineref
\input{references}
\else
\bibliography{ms}
\fi

\ifextended
\newpage
\appendix
\section{Notes on the Use of Clingo}\label{sec:reactive-clingo}

\leanparagraph{Reactive features} We established techniques
that allow for incrementally updating a program $\enc{P}_k$ for time or
count increment, where Alg.~\ref{alg:increment-tick} identifies at each
tick new rules $G^+$ that have to be added to the previous translation,
and expired ones $G^-$ that must be deleted.

In search of existing systems that might allow such incremental program
update, we considered the state-of-the-art ASP solver
Clingo~\cite{clingo14}, which comes with an API for reactive/multi-shot
solving.\footnote{Clingo 5.1.0. API:
  \url{https://potassco.org/clingo/python-api/current/clingo.html}}
These functionalities are based on~\cite{reactive11}, have since
evolved~\cite{asp-for-stream-reasoning-gebser12,clingo14} and
successfully applied; e.g. viz.~\cite{ricochet}. Unfortunately, for our
purposes, control features in Clingo are not applicable.

First, the control features in Clingo allow addition of new rules, but
not removal of existing ones. Technically, removing might be simulated
by setting a designated switch atom to false. However, this approach
would imply that the program grows over time. Second, we considered
using reactive features as illustrated for Rule $r$ of
Ex.~\ref{ex:alpha-high-abstraction}, using a program part that is
parameterized for stream variables, including that of tick $(t,c)$.

\vspace*{1ex}
\begin{footnotesize}  
\noindent\texttt{\#program tick(t, c, v).\\[-1ex]
\#external now(t).\\[-1ex]
\#external cnt(c).\\[-1ex]
\#external alpha\_at(v,t).\\[-1ex]
high\_at(t) :- w\_time\_2\_alpha(v,t), t >= 18.\\[-1ex]
w\_time\_2\_alpha(v,t) :- now(t), alpha\_at(v,t).\\[-1ex]
w\_time\_2\_alpha(v,t) :- now(t), alpha\_at(v,t-1).\\[-1ex]
w\_time\_2\_alpha(v,t) :- now(t), alpha\_at(v,t-2).} 
\end{footnotesize}
\vspace*{1ex}

\noindent However, this encoding is not applicable, since atoms in rule
heads cannot be redefined, i.e., they cannot be grounded more than once.

\leanparagraph{Model update}
For stratified programs (which have a unique model),
repeatedly calling Clingo (by standard one-shot solving) on the encoded
program $\enc{P}$ is a practical solution. However, when a program has
multiple models, we then have no link between the output of successive
ticks, i.e., the model may arbitrarily change. For instance, consider program

\begin{center}
  {\small \texttt{a
      :- \mbox{not b,} \mbox{not c.~} b :- \mbox{not a,} \mbox{not c.~} c :- \mbox{not a,} \mbox{not b.}}}
\end{center}
\noindent Using Clingo 5.1.0, the answer set of the program that is
returned first is {\small \texttt{\{a\}}}, which remains an answer set
if we add rule {\small \texttt{\mbox{a :- not c}}}. However, the first
reported answer set now is {\small \texttt{\{c\}}}.

\section{Proofs}

\subsubsection*{Proof for Lemma~\ref{lemma:time-window-underlies}}

Let $S=(T,\intpr)$ be a stream that underlies tick stream
$\dot{S}=(K,\tickIntpr)$, such that
$K=\langle (t_1,c_1),\dots,(t_m,c_m) \rangle$. By definition,
$T=[t_1,t_m]$ and
$\intpr(t)=\bigcup \{ \tickIntpr(t,c) \mid (t,c) \in K\}$ for all $t \in T$.
We recall that $\wfnTime_n(S)$ (resp. $\wfnTime_n(\dot{S})$) abbreviates
$\wfnTime_n(S,t_m)$ (resp. $\wfnTime_n(\dot{S},(t_m,c_m)$).
Thus, by definition, $\wfnTime_n(\dot{S})=(K',\tickIntpr|_{K'})$, where
$K'=\{(t',c') \in K \mid \max\{t_1,t-n\} \leq t' \leq t\}$, and
$\wfnTime_n(S)=(T',\intpr|_{T'})$, where $T'=[t',t_m]$ and
$t'=\max\{t_1,t-n\}$. We observe that $t'$ is the minimal time point
selected also in $K'$, i.e.,
$K'=\langle (t_k,c_k),\dots,(t_m,c_m) \rangle$ implies $t_k=t'$. It
remains to show that
$(\intpr|_{T'})(t) = \bigcup \{ (\tickIntpr|_{K'})(t,c) \mid (t,c) \in K'\}$
for all $t \in T'$. This is seen from the fact that neither
$\wfnTime_n(S)$ nor $\wfnTime_n(\dot{S})$ drops any data within $T'$.
We conclude that $\wfnTime_n(S)$ underlies $\wfnTime_n(\dot{S})$.

\subsubsection*{Proof Sketch for Lemma~\ref{lemma:tuple-window-underlies}}

The argument is similar as for
Lemma~\ref{lemma:time-window-underlies}. The central observation is that
a tick stream provides a more fine-grained control over the information
available in streams by introducing an order on tuples in addition to
the temporal order. Each time point in a stream is assigned a set of
atoms, whereas each tick in a tick stream is assigned at most one
atom. The tuple-based window function~$\wfnTuple_n$ always counts atoms
backwards (from right end to left) and then selects the timeline
$[t_1,t]$ with the latest possible left time point $t_1$ required to
capture $n$ atoms. While for tick streams, the order is unique, but
multiple options exist for streams in general. If the tuple window
$\wfnTuple_n(S)$ is based on the order in which atoms appeared in $S$,
then it selects the same atoms as $\wfnTuple_n(\dot{S})$, and thus the
same timeline. Consequently, $\wfnTuple_n(S)$ underlies
$\wfnTuple_n(\dot{S})$.

\subsubsection*{Proof Sketch for Proposition~\ref{prop:soundness-static-encoding}}

The desired correspondence is based on two translations: a LARS program
$P$ (at a time $t$) into a logic program $\enc{P}=\mi{LarsToAsp}(P,t)$
(due to Algorithm~\ref{alg:lars-program-to-asp}), and the encoding of a
stream $S$ as set $\enc{S}$ of atoms.
Given a fixed timeline $T$, we may view a stream
$S=(T,\intpr)$ as a set of pairs
$\{ (a(\vec{x}),t) \mid a(\vec{x}) \in \intpr(t), t \in T \}$. This is
the essence of a stream encoding $\enc{S}$ for the tick stream
$\dot{S}=(K,\tickIntpr)$; $\enc{S}$ includes the analogous time-pinned
atoms:
$\{ \timePinned{a}(\vec{x},t) \mid a(\vec{x}) \in \tickIntpr(t,c), (t,c)
\in K\}$. With respect to the correspondence, atoms of form
$\tickPinned{a}(\vec{x},t,c)$, $\cnt(c)$ and $\tick(t,c)$ in $\enc{S}$
can be considered auxiliary, as well as the specific counts used in the
tick pattern $K$ to obtain time-pinned atoms
$\timePinned{a}(\vec{x},t)$. Counts play a role only for the specific
selection of tuple-based windows, which are assumed to reflect the order
of the tick stream. Thus, we may view a stream encoding~$\enc{S}$
essentially as a different representation of stream $S$; additional atoms
can be abstracted away as they have no correspondence in the original
LARS stream or program. We thus consider only the time-pinned atoms in
an encoded stream to read off a LARS stream.

Thus, it remains to argue the soundness of the transformation
$\mi{LarsToAsp}$, which returns a program of form
$Q \cup R \cup \{\now(t)\}$, where $\now(t)$ is auxiliary. The set $Q$
simply identifies time-pinned atoms $\timePinned{a}(\vec{X},\dot{N})$
with $a(\vec{X})$ in case $\dot{N}$ is the current time point. This is
the information provided by predicate $\now$ for which a unique atom
exists. Thus, $Q$ ensures that a time-pinned atom
$\timePinned{a}(\vec{x},t)$ is available if $a(\vec{x},t)$ is derived,
and vice versa; $Q$ thereby only accounts for redundant representations
of atoms that currently hold.

Towards $R$, we get the translation by the function
$\mi{larsToAspRules}$ which returns a set of encoded rules for every
LARS rule $r$. First, the $\mi{baseRule}$ is the corresponding ASP rule,
which introduces a new symbol $\mi{atm}(e)$ for every extended atom in
the rule that is not an ordinary atom. In order to ensure that the base
rule $\enc{r}$ fires in an interpretation just if the original rule $r$
fires in the corresponding interpretation of program $P$, for each body
element $\mi{atm}(e)$ in $\enc{r}$
the set of rules to derive $\mi{atm}(e)$ in lines (14)-(21) is provided;
the correspondence between
$@_Ta(\vec{X})$ and $\timePinned{a}(\vec{X},T)$ is already given by
construction. Thus, each interpretation stream $I \supseteq D$ for
$P$ has a corresponding
interpretation $\enc{I}$ for $\mi{LarsToAsp}(P)$ in which 
besides the time-pinned atoms the atoms $\mi{atm}(e)$ and
$\mi{spoil}_e(\vec{X})$ occur depending on
support from (i.e., firing) of the rules in (14)-(21), such that they correctly
reflect the value of the window atoms $e$ in $I$.

As each atom in an answer of an ordinary ASP program must 
derived by a rule, it is not hard to see that
every answer set of $\enc{P} =
\mi{LarsToAsp}(P,t) \cup \enc{D}$ is of the form $\enc{I}$, where $I
\supseteq D$ is an interpretation stream for $D$. We thus need to show the following: $I \in \AS(P,D,t)$ holds iff $\enc{I}$ is an answer set of $\enc{P}$. We do this for ground $P$ (the extension to non-ground $P$ is straightforward).

($\Rightarrow$) For the only-if direction, we show that if $I \in
\AS(P,D,t)$, that is, $I$ is a minimal model for
the reduct $P^{M,t}$ where $M=\tup{I,W,B}$, then (i) $\enc{I}$ is a model of the reduct $\enc{P}^{\enc{I}}$,
and  (ii) no interpretation $J'\subset \enc{I}$  is a model of $\enc{P}^{\enc{I}}$. 
As for (i), we can concentrate by construction of $\enc{I}$ on the 
base rules $\enc{r}=\mi{baseRule}(r)$ in $\enc{P}^{\enc{I}}$ (all other rules will be satisfied). If $\enc{I}$
satisfies $\body(\enc{r})$, then by construction $I$
satisfies $\body(r)$; as $I$ is a model of $P^{M,t}$, it follows that
$I$ satisfies $\head(r)$; but then, by construction, $\enc{I}$ satisfies
$\head(\enc{r})$.
As for (ii), we assume towards a contradiction that some
$J'\subset\enc{I}$ satisfies $\enc{P}^{\enc{I}}$. We then consider the
stream $J \supseteq D$ that is induced by $J'$, and any rule $r$ in the
reduct $P^{M,t}$. If $J$ does not satisfy $\body(r)$, then $J$ satisfies
$r$; otherwise, if $J$ satisfies $\body(r)$, then as $\enc{r}$ is 
in the reduct $\enc{P}^{\enc{I}}$, we have that $\enc{I}$ falsifies
each atom $\mi{atm}(e)$ in $\body^-(\enc{r})$, and as $J'\subset \enc{I}$,
also $J'$ falsifies each such $\mi{atm}(e)$. Furthermore, as $J$
satisfies each atom $e\in\body^+(r)$, from the rules for
$\mi{atm}(e)$ among (14)-(21) in the reduct $\enc{P}^{\enc{I}}$ we obtain that $J'$
satisfies each atom $\mi{atm}(e)$ in $\body^+(\enc{r})$. That
is, $J'$ satisfies $\body(\enc{r})$. As $J'$ satisfies $\enc{r}$, we
then obtain that $J'$ satisfies $\head(\enc{r})$. The latter
means that $J$ satisfies $\head(r)$, and thus $J$ satisfies $r$. As $r$
was arbitrary from the reduct $P^{M,t}$, we obtain that $J \subset I$ is
a model of $P^{M,t}$; this however contradicts that $I$ is a minimal
model of $P^{M,t}$, and thus (ii) holds.

($\Leftarrow$) For the if direction, we argue similarly. Consider an
answer set $\enc{I}$ of $\enc{P}$. To show that $I \in \AS(P,D,t)$, we
establish that (i) $I$ is a model of $P^{M,t}$ and (ii) no model
$J\subset I$ of $P^{M,t}$ exists.  As for (i), since in $\enc{I}$ the
atoms $\mi{atm}(e)$ correctly reflect the value of the window atoms $e$
in $I$, for each $r$ in $P^{M,t}$ the rule $\enc{r}=\mi{baseRule}(r)$ is
in $\enc{P}^{\enc{I}}$; as $\enc{I}$ satisfies $\enc{r}$, we conclude
that $I$ satisfies $r$. As for (ii), we show that every model $J$ of
$P^{M,t}$ must contain $I$, which then proves the result.

To establish this, we use the fact that $\enc{I}$ can be generated by a
sequence $\rho = r_1,r_2,r_3\ldots,r_k$ of rules from $\enc{P}^{\enc{I}}$ with
distinct heads such that (a) $\enc{I} = \{\head(r_1), \ldots\head(r_k)\} =: \enc{I}_k$ and
(b) $\enc{I}_{i-1} = \{\head(r_1),\ldots,\head(r_{i-1})\}$  satisfies
$\body^+(r_i)$, for every $i=1,\ldots,k$.

In that, we use the assertion that no 
cyclic positive dependencies through time-based window atoms
$\window^n\Box a$ occur. Formally, positive dependency is defined as
follows: an atom $@_{t_1}b$ positively depends on an
atom $@_{t_2}a$ in a ground program $P$ at $t$, if some rule
$r\in P$ exists with $\head(r) = @_{t_1}b $ and such that either (a)
$@_{t_2}a \in \body^+(r)$, or (b) $\window^n @_{t_2}a \in
\body^+(r)$ or (c) $\window^n\star a \in \body^+(r)$, $\star \in
\{\Box, \Diamond\}$, where in (b) and (c) 
$t_2 \in [t-n,t]$ holds.
As in $\mi{LarsToAsp}(P,t)$, all ordinary atoms $a$ are here viewed as $@_t a$.
A cyclic positive dependency through $\window^n\Box a$ is then 
a sequence $@_{t_0} a_0$, $@_{t_1} a_1$, \ldots, $@_{t_k} a_k$, $k\geq
1$, such that $@_{t_i} a_i$ positively depends on ${@_{t_{(i+1)\mod k}}\, a_{(i+1)\mod k}}$,
for all $i=0,\ldots,k$ and $a_0=b$ and $a_1=a$ for case (c) with $\star=\Box$.

Given that no positive cyclic dependencies through atoms $\window^n\Box
a$ occur in $P$ at $t$, and thus in $P^{M,t}$, 
we can w.l.o.g.\ assume that whenever $r_i$ in $\rho$ has a head $\encWindowAtom_e$ for
a window atom $e=\window^n\Box a$, each rule $r_j$ in $\rho$ with a head
$a_@(t')$, where 
$t' \in [t-n,t]$,
 precedes $r_i$,
i.e., $j<i$ holds.

By induction on $i\geq 1$, we can now show that if $\head(r_i)=\mi{atm}(e)$, then every model $J$ of $P^{M,t}$ must satisfy $e$;
consequently, at $i=k$, $J$ must contain $I$. From the form of the rules
$\mi{baseRule}(r)$ and $\mi{windowRules}(e)$, the correspondence
between $\enc{P}^{\enc{I}}$ and $P^{M,t}$, and the fact that the external data 
are facts, only the case $e=\window^n\Box a(\vec{X})$ needs a further argument. 
Now if $r_i$ is the rule $\encWindowAtom_e \leftarrow a(\vec{X}),\naf\, \mi{spoil}_e(\vec{X})$
on line (16), then $\enc{I}$ must satisfy $a$ and falsify
$\mi{spoil}_e(\vec{X})$; in turn, every $a_@(t',\vec{X})$  must be true in $\enc{I}$, for
$t' \in [t-n,t]$.
From the induction hypothesis, we obtain that
$@_{t'}a(\vec{X})$ is true in every model $J$ of $P^{M,t}$,
$t' \in [t-n,t]$,
and thus $e=\window^n\Box a(\vec{X})$ is true as well. This proves the claim
and concludes the proof of the if-case, which in turn establishes the
claimed correspondence between $\AS(P,D,t)$ and
the answer sets of $\enc{P} =\mi{LarsToAsp}(P,t) \cup \enc{D}$.

\medskip

\noindent{\it Remark.} The condition on cyclic positive dependencies
excludes that rules $b \leftarrow \window^n\Box a$ and $a \leftarrow b$
occur jointly in a program. A stricter notion of dependency that allows for
co-occurrence is to request in (c) for $\star=\Box$ in addition $t_2<t$; then
e.g.\ any LARS program where the rule heads are ordinary atoms
is allowed, and Proposition~\ref{prop:soundness-static-encoding} remains valid. 

\subsubsection*{Proof Sketch for Proposition~\ref{prop:cut-off}}

Assume a LARS program $P$ and two tick data streams $D=(K,\tickIntpr)$
and $D'=(K',\tickIntpr')$ at tick $(t_m,c_m)$ such that $D' \subseteq D$
and $K'=\langle (t_k,c_k),\dots,(t_m,c_m)\rangle$. Furthermore, assume
that (*) all atoms/time points accessible from any window in $P$ are
included in $D'$.
We want to show
$\AS^\intensional(\enc{P}_{D,m}) =
\AS^\intensional(\enc{P}_{D',m})$. The central observation is that rules
need to fire in order for intensional atoms to be included in the answer
set, and that no rules can fire based on outdated ticks. Thus, these
ticks can also be dropped.

In more detail, we assume
$\AS^\intensional(\enc{P}_{D,m}) \neq \AS^\intensional(\enc{P}_{D',m})$
towards a contradiction. That is to say, a difference in evaluation
arises based on data in $D \setminus D'$, i.e., atoms appearing before
tick $(t_k,c_k)$. Consider any extended atom $e$ of a (LARS) rule
$r \in P$, where the body holds only for one of the two encodings (in
the same partial interpretation). Due to the assumption (*), we can
exclude a difference arising from a window atom of form
$\window^w \star a$, $\star \in \{\Diamond,\Box,@_T\}$.

If $e$ is an atom $a$, it holds in $\enc{P}_{D,m}$ iff it holds in
$\enc{P}_{D',m}$ since an ordinary atom in the answer set of the
encoding corresponds to an atom holding at the current time point, and
both $D$ and $D'$ include the current time point.

The last option is $e = @_T a$, which may reach back beyond $(t_k,c_k)$
but is viewed in the incremental encoding as syntactic shortcut for
$\timeWindow{\infty} @_T a$. That is, in this case we have $D'=D$ and
thus the encodings coincide.

We conclude that assuming
$\AS^\intensional(\enc{P}_{D,m}) \neq \AS^\intensional(\enc{P}_{D',m})$
is contradictory due to these observations. Spelling out the details
fully involves essentially a case distinction on the incremental window
encodings and arguing about the relationship between $(t_k,c_k)$, the
respective expiration annotations, and the fact that rules accessing
atoms at ticks before $(t_k,c_k)$ are have already expired.

\subsubsection*{Proof Sketch for Proposition~\ref{prop:equiv-static-and-incremental-encoding}}

\newcommand{\staticP}{\ensuremath{\enc{P} \cup \enc{D}}\xspace}
\newcommand{\incrP}{\ensuremath{\enc{P}_{D,m}}\xspace}

We argue based on the commonalities and differences of the static
encoding \staticP and the incremental encoding \incrP.
%
Instead of body predicates $\now(\dot{N})$ and $\cnt(\dot{C})$, that are
instantiated in \staticP due to the predicates $\now(t)$
and $\cnt(c)$, \incrP directly uses the instantiations of tick
variables.
%
In both encodings, the window atom is associated with a set of rules
that needs to model the temporal quantifier ($\Diamond$,$\Box$,$@_t$) in
the correct range of ticks as expressed by the LARS window atom. This
window always includes the last tick. While \staticP is based on a
complete definition how far the window extends, \incrP updates this
definition tick by tick. In particular, the oldest tick that is not
covered by the window anymore corresponds to the expiration annotation
in \incrP.

The case $\timeWindow{n} \Diamond a(\vec{X})$ is as follows: in the static rule encoding,
\begin{displaymath}
  \encWindowAtom_e(\vec{X}) \leftarrow \mi{now}(\dot{N}),
  \timePinned{a}(\vec{X},T)\,,
\end{displaymath}
given $\now(t)$, time variable $T$ will be grounded with 
$t-n,\dots,t-0$. That is, we get a set of rules
\begin{displaymath}
  \begin{array}{lrcl}
  (r_{0}) & \encWindowAtom_e(\vec{X}) & \leftarrow & \mi{now}(t), \timePinned{a}(\vec{X},t)\\
   & & \vdots\nonumber\\
    (r_{n}) & \encWindowAtom_e(\vec{X}) & \leftarrow & \mi{now}(t), \timePinned{a}(\vec{X},t-n)\,,
  \end{array}
\end{displaymath}
where arguments $\vec{X}$ will be grounded due to data and inferences.
We observe that $(r_0)$ is the rule that is inserted to the incremental
program \incrP at time $t$ (minus predicate $\now(t)$, since in \incrP
variable~$T$ is instantiated directly with $t$ to obtain
$\timePinned{a}(\vec{X},t)$), and all rules up to $r_n$ remain from
previous calls to $\mi{IncrementalRules}$. Rule $r_n$ will expire at
$t+1$, i.e., the exact time when it will not be included in
$\enc{P} \cup \enc{D}$ anymore. The cases for
$\timeWindow{n}@_T a(\vec{X}), \timeWindow{n}\Box a(\vec{X}),
\tupleWindow{n}\Diamond a(\vec{X})$ and $\tupleWindow{n}@_T a(\vec{X})$
are analogous; the remaining case $\tupleWindow{n}\Box a(\vec{X})$ has
been argued earlier.

Finally, $\enc{P}_{D,m}$ includes a stream encoding, which is also
incrementally maintained: at each tick $(t,c)$ the tick atom
$\tick(t,c)$ is added, and in case of a count increment, the time-pinned
atom $\timePinned{a}(\vec{X},t)$ and the tick-pinned atoms
$\tickPinned{a}(\vec{X},t,c)$ are added to \incrP as in $\enc{D}$. This
way, we have a full correspondence with the static stream encoding
$\enc{D}$.

Thus, at every tick $(t,c)$, \staticP and \incrP have the same data and
express the same evaluations. Disregarding auxiliary atoms, we conclude
that their answer sets coincide.

\subsubsection*{Proof Sketch for Theorem~\ref{theorem:incremental-lars-evaluation}}

Given a LARS program $P$, a tick data
  stream  $D=(K,\tickIntpr)$ at tick $(t,c)$ by Prop.~\ref{prop:soundness-static-encoding} 
$S$ is an answer stream of $P$ for $D$ at $t$ iff $\enc{S}$ is an answer set of
$\enc{P} \cup \enc{D}$, where $\enc{P}=\mi{LarsToAsp(P,t)}$. By
Prop.~\ref{prop:equiv-static-and-incremental-encoding}, for any set
$X$ we have that $X \cup \{\now(t),\cnt(c)\}$ is an answer set of
$\enc{P} \cup \enc{D}$ iff $X$ is an answer set of $\enc{P}_{D,m}$
(modulo auxiliary atoms). In particular this holds for $X =
\enc{S}$. As $\{\now(t),\cnt(c)\}\subseteq \enc{S}$, we obtain that 
$S$ is an answer stream of $P$ for $D$ at $t$ iff $\enc{S}$ is an answer
set of $\enc{P}_{D,m}$, which is the result.

\section{Details of Evaluation Results}
(See pages~\pageref{tab:tp:caching:A1}--\pageref{tab:winsize:content:B2}.)
\clearpage

\newcommand{\tablePos}{t}

\begin{table}[\tablePos]
\caption{Results for \emph{A1}. Variable window size $n$. Results for 1000 timepoints in seconds.}
\label{tab:tp:caching:A1}
\begin{footnotesize}
\begin{center}
\begin{tabular}{|c|*{3}{c}|*{3}{c}|} 
& \multicolumn{3}{c|}{Clingo} & \multicolumn{3}{c|}{Incremental} \\
 $n$ & $\tTotal$ & $\tInit$ &  $\tTick$ &  $\tTotal$ & $\tInit$  & $\tTick$ \\[1.0ex]
20 & 14.296 & 0.017 & 0.014 &  2.638 & 0.016 & 0.002\\
40 & 20.526 & 0.018 & 0.02 &   3.006 & 0.018 & 0.002\\
80 & 34.491 & 0.025 & 0.034 &  2.938 & 0.018 & 0.002\\ 
120 & 49.249 & 0.027 & 0.049 & 3.439 & 0.019 & 0.003\\
160 & 64.661 & 0.028 & 0.064 & 3.554 & 0.017 & 0.003\\
200 & 79.105 & 0.036 & 0.079 & 3.674 & 0.018 & 0.003\\ 
\end{tabular}
\end{center}
\end{footnotesize}
\end{table}

\begin{table}[\tablePos]
\centering
\caption{Results for \emph{A2}. Variable window size $n$. Runtime for 1000 timepoints in seconds.}
\label{tab:tp:caching:A2}
\begin{footnotesize}
\begin{center}
\begin{tabular}{|c|*{3}{c}|*{3}{c}|} 
& \multicolumn{3}{c|}{Clingo} & \multicolumn{3}{c|}{Incremental} \\
 $n$ & $\tTotal$ & $\tInit$ &  $\tTick$ &  $\tTotal$ & $\tInit$  & $\tTick$ \\[1.0ex]
 20 & 15.259 & 0.02 & 0.015 & 2.869 & 0.016 & 0.002\\
40 & 23.123 & 0.02 & 0.023  & 3.201 & 0.018 & 0.003\\ 
80 & 35.962 & 0.022 & 0.035 & 3.365 & 0.019 & 0.003\\
120 & 49.068 & 0.026 & 0.049 & 3.547 & 0.02 & 0.003\\
160 & 61.983 & 0.03 & 0.061  & 3.842 & 0.018 & 0.003\\
200 & 80.899 & 0.036 & 0.08  & 3.7 & 0.019 & 0.003 \\ 
\end{tabular}
\end{center}
\end{footnotesize}
\end{table}

\begin{table}[\tablePos]
\centering
\caption{Results for \emph{A1}. Variable timepoints $\mi{tp}$. Runtime for window size $n=60$ in seconds.}
\label{tab:winsize:caching:A1}
\begin{footnotesize}
\begin{center}
\begin{tabular}{|c|*{3}{c}|*{3}{c}|} 
& \multicolumn{3}{c|}{Clingo} & \multicolumn{3}{c|}{Incremental} \\
 $\mi{tp}$ & $\tTotal$ & $\tInit$ &  $\tTick$ &  $\tTotal$ & $\tInit$  & $\tTick$ \\[1.0ex]
100 & 2.78 & 0.026 & 0.027    & 0.368 & 0.023 & 0.003\\
200 & 5.49 & 0.022 & 0.027   & 0.674 & 0.02 & 0.003\\
300 & 8.269 & 0.022 & 0.027  & 1.072 & 0.026 & 0.003\\
400 & 11.379 & 0.026 & 0.028 & 1.307 & 0.02 & 0.003\\
500 & 14.192 & 0.024 & 0.028 & 1.695 & 0.017 & 0.003\\
600 & 16.709 & 0.023 & 0.027 & 1.945 & 0.02 & 0.003\\
700 & 20.049 & 0.021 & 0.028 & 2.217 & 0.017 & 0.003\\
800 & 22.534 & 0.021 & 0.028 & 2.627 & 0.018 & 0.003\\
900 & 25.892 & 0.024 & 0.028 & 3.183 & 0.022 & 0.003\\
1000 & 27.501 & 0.021 & 0.027  & 3.42 & 0.021 & 0.003\\ 
\end{tabular}
\end{center}
\end{footnotesize}
\end{table}

\begin{table}[\tablePos]
\centering
\caption{Results for \emph{A2}. Variable timepoints $\mi{tp}$. Runtime for window size $n=60$ in seconds.}
\label{tab:winsize:caching:A2}
\begin{footnotesize}
\begin{center}
\begin{tabular}{|c|*{3}{c}|*{3}{c}|} 
& \multicolumn{3}{c|}{Clingo} & \multicolumn{3}{c|}{Incremental} \\
 $\mi{tp}$ & $\tTotal$ & $\tInit$ &  $\tTick$ &  $\tTotal$ & $\tInit$  & $\tTick$ \\[1.0ex]
 100 & 2.998 & 0.026 & 0.029 & 0.418 & 0.019 & 0.003\\
200 & 5.727 & 0.023 & 0.028 & 0.89 & 0.017 & 0.004\\
300 & 9.06 & 0.026 & 0.03   & 1.097 & 0.021 & 0.003\\
400 & 11.783 & 0.021 & 0.029 & 1.563 & 0.02 & 0.003\\
500 & 14.26 & 0.021 & 0.028 & 1.81 & 0.017 & 0.003\\
600 & 17.439 & 0.02 & 0.029 & 2.181 & 0.021 & 0.003\\
700 & 20.321 & 0.021 & 0.028 & 2.438 & 0.018 & 0.003\\
800 & 23.3 & 0.02 & 0.029   & 3.371 & 0.02 & 0.004\\
900 & 26.51 & 0.021 & 0.029 & 3.22 & 0.018 & 0.003\\
1000 & 30.077 & 0.024 & 0.03 & 3.5 & 0.019 & 0.003\\ 
\end{tabular}
\end{center}
\end{footnotesize}
\end{table}

\begin{table}[\tablePos]
  \caption{Results for \emph{B1}. Variable window size $n$. Runtime in
    seconds for 1000 timepoints.}
\label{tab:tp:content:B1}
\begin{footnotesize}
\begin{center}
\begin{tabular}{|c|*{3}{c}|*{3}{c}|}
& \multicolumn{3}{c|}{Clingo} & \multicolumn{3}{c|}{Incremental} \\
 $n$ & $\tTotal$ & $\tInit$ &  $\tTick$ &  $\tTotal$ & $\tInit$  & $\tTick$ \\[1.0ex]
 20 & 26.158 & 0.018 & 0.026 & 15.641 & 0.292 & 0.015 \\
 40 & 55.898 & 0.021 & 0.055 & 16.726 & 0.315 & 0.016 \\
 80 & 425.853 & 0.019 & 0.425 & 21.135 & 0.299 & 0.02 \\
 120 & - & - & - & 25.909 & 0.304 & 0.025 \\
 160 & - & - & - & 30.659 & 0.363 & 0.03 \\
 200 & - & - & - & 33.541 & 0.306 & 0.033\\
\end{tabular}
\end{center}
\end{footnotesize}
\end{table}

\begin{table}[\tablePos]
\caption{Results for \emph{B2}. Variable window size $n$. Runtime for 1000 timepoints in seconds.}
\label{tab:tp:content:B2}
\begin{footnotesize}
\begin{center}
\begin{tabular}{|c|*{3}{c}|*{3}{c}|} 
& \multicolumn{3}{c|}{Clingo} & \multicolumn{3}{c|}{Incremental} \\
 $n$ & $\tTotal$ & $\tInit$ &  $\tTick$ &  $\tTotal$ & $\tInit$  & $\tTick$ \\ 
 20 & 24.138 & 0.018 & 0.024 &  34.717 & 0.292 & 0.033 \\
 40 & 38.478 & 0.019 & 0.038 &  35.744 & 0.333 & 0.034 \\
 80 & 71.827 & 0.024 & 0.071 &  25.767 & 0.298 & 0.025 \\
 120 & 104.723 & 0.023 & 0.104 & 33.788 & 0.29 & 0.033 \\
 160 & 148.257 & 0.031 & 0.148 & 31.1 & 0.303 & 0.03 \\
 200 & 181.991 & 0.028 & 0.181 & 37.612 & 0.33 & 0.036 \\ 
\end{tabular}
\end{center}
\end{footnotesize}
\end{table}

\begin{table}[\tablePos]
\centering
\caption{Results for \emph{B1}. Variable timepoints $\mi{tp}$. Window size $n=60$ in seconds.}
\label{tab:winsize:content:B1}
\begin{footnotesize}
\begin{center}
\begin{tabular}{|c|*{3}{c}|*{3}{c}|} 
& \multicolumn{3}{c|}{Clingo} & \multicolumn{3}{c|}{Incremental} \\
 $\mi{tp}$ & $\tTotal$ & $\tInit$ &  $\tTick$ &  $\tTotal$ & $\tInit$  & $\tTick$ \\[1.0ex]
100 &   8.57 & 0.026 & 0.085 & 1.895 & 0.32 & 0.015 \\
200 & 16.392 & 0.022 & 0.081 & 3.971 & 0.293 & 0.018 \\
300 & 31.568 & 0.022 & 0.105 & 6.82 & 0.475 & 0.021 \\
400 & 40.927 & 0.025 & 0.102 & 8.518 & 0.332 & 0.02 \\
500 &  55.313 & 0.021 & 0.11 & 10.64 & 0.351 & 0.02 \\
600 & 69.548 & 0.021 & 0.115 & 12.816 & 0.353 & 0.02 \\
700 &   -     &  -     &  -     & 15.773 & 0.333 & 0.021 \\
800&   -     &   -    &    -   & 16.756 & 0.318 & 0.02 \\
900&    -    &   -    &    -   & 16.96 & 0.298 & 0.018 \\
1000&   -     &   -    &   -   & 18.602 & 0.298 & 0.018 \\ 
\end{tabular}
\end{center}
\end{footnotesize}
\end{table}

\begin{table}[\tablePos]
\centering
\caption{Results for \emph{B2}. Variable timepoints $\mi{tp}$. Window size $n=60$ in seconds.}
\label{tab:winsize:content:B2}
\begin{footnotesize}
\begin{center}
\begin{tabular}{|c|*{3}{c}|*{3}{c}|} 
& \multicolumn{3}{c|}{Clingo} & \multicolumn{3}{c|}{Incremental} \\
 $\mi{tp}$ & $\tTotal$ & $\tInit$ &  $\tTick$ &  $\tTotal$ & $\tInit$  & $\tTick$ \\[1.0ex]
100 & 4.974  & 0.029 & 0.049& 1.838 & 0.299 & 0.015 \\
200 & 10.06  & 0.021 & 0.05 & 3.982 & 0.304 & 0.018 \\
300 & 15.023 & 0.02  & 0.049& 6.126 & 0.359 & 0.019 \\
400 & 20.574 & 0.019 & 0.051& 9.062 & 0.29 & 0.021 \\
500 & 26.075 & 0.02  & 0.052& 11.625 & 0.289 & 0.022 \\
600 & 31.68  & 0.02  & 0.052& 14.974 & 0.297 & 0.024 \\
700 & 36.35  & 0.02  & 0.051& 18.301 & 0.29 & 0.025 \\
800 & 42.391 & 0.021 & 0.052& 22.947 & 0.286 & 0.028 \\
900 & 48.254 & 0.021 & 0.053& 28.979 & 0.366 & 0.031 \\
1000 & 54.35 & 0.02  & 0.054 & 28.993 & 0.334 & 0.028 \\ 
\end{tabular}
\end{center}
\end{footnotesize}
\end{table}


\fi

\label{lastpage}
\end{document}